\documentclass[epj]{svjour}
\usepackage{amsmath}
\usepackage{amssymb}
\usepackage{graphicx}
\usepackage{dcolumn}
\usepackage{bm}

\def\d{\mathrm{d}}
\def\vec#1{\ensuremath{\mathbf{#1}}}

\begin{document}

\title{Fluctuation spectra of free and supported membrane pairs} 

\author{Rolf-J\"urgen Merath\inst{1,}\inst{2} \and Udo Seifert\inst{3}}



\institute{   
   Max-Planck-Institut f\"ur Metallforschung,
   Heisenbergstra\ss e 3,
   70569 Stuttgart,
   Germany
\and 
    Institut f\"ur Theoretische und Angewandte Physik, 
    Universit\"at Stuttgart, 
    Pfaffenwaldring 57, 
    70569 Stuttgart, 
    Germany  
\and 
   II.\ Institut f\"ur Theoretische Physik,
   Universit\"at Stuttgart,
   Pfaffenwaldring 57,
   70550 Stuttgart,
   Germany
}

\date{Received: date / Revised version: date}
\date{\today} 

\abstract{
Fluctuation spectra of fluid compound membrane systems are calculated. 
The systems addressed contain two (or more) almost parallel membranes that are connected by harmonic tethers or by 
a continuous, harmonic confining potential. Additionally, such a compound system can be attached to a supporting substrate. 
We compare quasi-analytical results for tethers with analytical results for corresponding continuous models 
and investigate under what circumstances the discrete nature of the tethers
actually influences the fluctuations. 
A tethered, supported membrane pair with similar bending rigidities and stiff tethers 
can possess a nonmonotonic fluctuation spectrum with a maximum. 
A nonmonotonic spectrum with a maxi\-mum and a minimum can occur 
for an either free or supported membrane pair of rather different bending rigidities and for stiff tethers.
Typical membrane displacements are calculated for supported membrane pairs with discrete or continuous interacting potentials.
Thereby an estimate of how close the constituent two membranes and the substrate typically approach each other is given. 
For a supported membrane pair with discrete or continuous interactions,  
the typical displacements of each membrane are altered with respect to a single supported membrane, 
where those of the membrane near the substrate are diminished and those of the membrane further away are enhanced. 
}

\authorrunning{R.-J.\ Merath and U.\ Seifert}

\maketitle

\section{Introduction}
\label{sec:1}
Research on biological membranes has benefitted considerably from studying well-defined
model systems~\cite{Lipowsky_Sackmann_Handbook_1995}. 
Membranes attached to a substrate can be investigated with surface-sensitive techniques
(e.\,g., reflection interference contrast microscopy, 
fluorescence micro\-scopy,
atomic force microscopy, surface plasmon resonance, and quartz crystal microbalance).
However, for solid-supported membranes, the substrate tends to demobilize and denaturalize membrane proteins. 
Hence, systems with a biocompatible surface are intended to 
mimick the biological membrane in its natural environment~\cite{Tanaka_Sackmann_Nature_2005_and_TrendsBT_2000}.
A promising and meanwhile widely used approach is to tether the membrane to the substrate, 
ensuring a larger distance between the membrane and the solid while keeping the membrane confined. 
Recent experimental achievements and developments permit 
(i) to micro- and nano-struc\-ture substrates in order to configure the pattern of attachment sites 
\cite{Arnold_Spatz_ChemPhysChem_2004,Roos_Spatz_ChemPhysChem_2003,Glass_Spatz_AdvFunctMater_2003}, 
(ii) to tailor the length and stiffness of the tether molecules (or solid pillars) 
\cite{Atanasov_Koeper_BioconjugateChem_2006,Foertig_Tanaka_MacromolSymp_2004,Purrucker_Tanaka_ChemPhysChem_2004,Giess_Knoll_BiophysJ_2004}, 
and (iii) to adapt the linker chemistry to 
different substrates~\cite{Wagner_Tamm_BiophysJ_2000,Atanasov_Koeper_BiophysJ_2005,Naumann_and_Bunjes_2003_and_1997}. 
Different structures of composite membranes are generally accompanied by different thermal fluctuation spectra, 
which have previously been measured for red blood cells \cite{Zilker_JPhys_1987} 
and for a bilayer membrane interacting with a laterally homogeneous substrate~\cite{Raedler_PRE_1995}.

Theoretically thermal shape undulations of mem\-branes have been addressed for various model systems. 
Fluctuation spectra of supported membranes have been calculated for a single membrane connected to a fixed support.
This connection was established either by a con\-ti\-nu\-ous harmonic confinement \cite{Gov_Zilman_Safran_PRL_2003} 
or by tethers at discrete attachment 
sites~\cite{Gov_Safran_PRE_2004,Lin_Brown_BiophysJ_2004,Merath_Seifert_PRE_2006,Dubus_Fournier_2006}. 
For the latter case of localized interactions, nonmonotonic fluctuation spectra have been observed. 
The substrate structure has been considered in the context of 
adhering supported membranes~\cite{Andelman_1999_and_2001}.
The attachment of a membrane to an elastic meshwork has also been studied~\cite{Fournier_2004}. 
Nethertheless, so far, fluctuations of an underlying support have not yet been included quantitatively 
into the analysis of membrane fluctuations.

\begin{figure*}[!t]   
\hfil \hspace{-0.0cm} \includegraphics[width=13.3cm, clip]{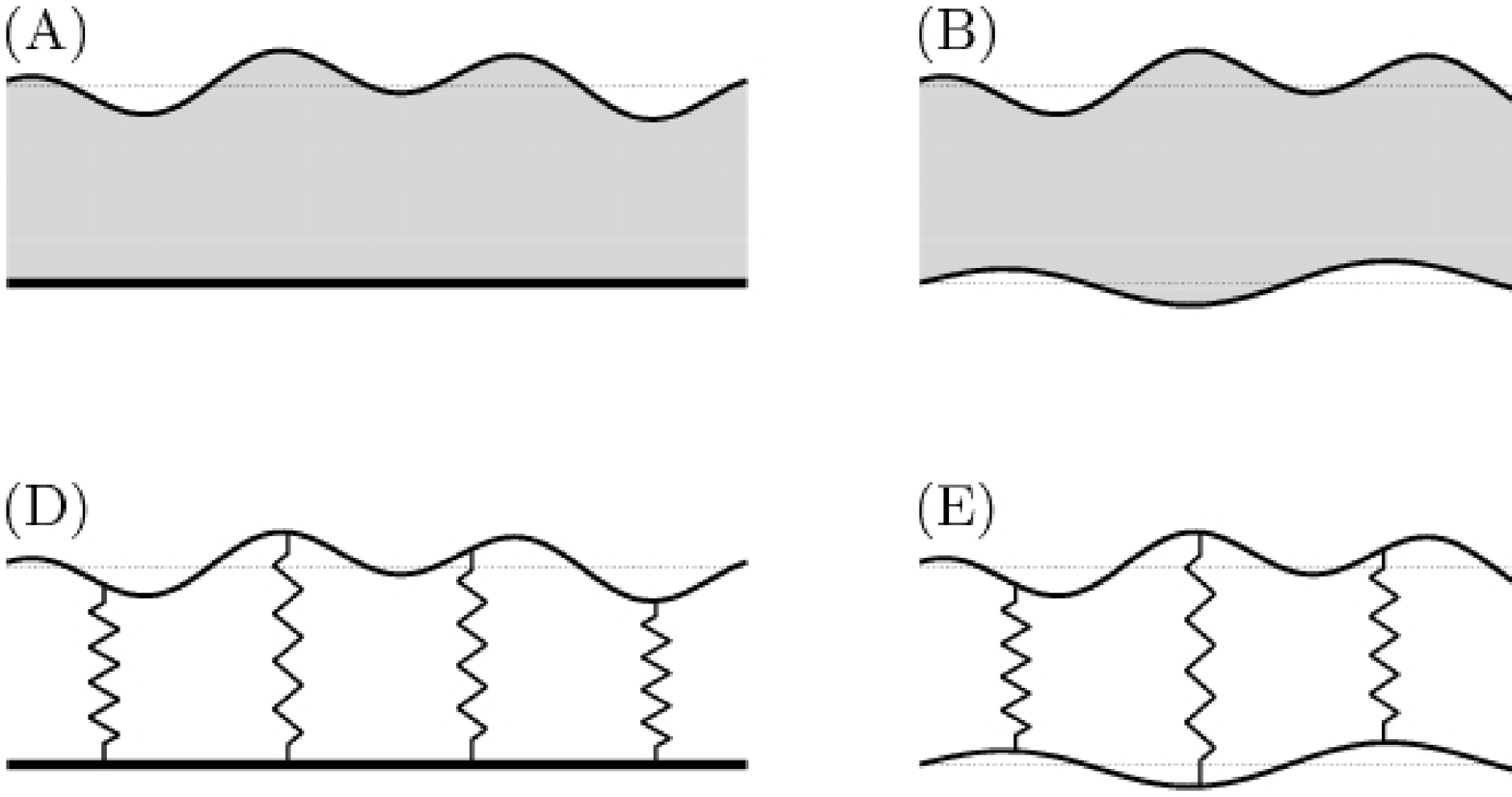} \hfil
\caption{Side-view sketches of the model systems addressed in this article. 
         These compound membrane systems are built up by almost parallel fluid membranes in an aqueous solution, potentially
         situated on a solid support (substrate). Temperature-driven shape fluctuations are influenced by the interactions between
         the membranes and/or the substrate. 
         The models A--C comprise a \emph{continuous} harmonic confining potential 
         between the constituent membranes and --- if present --- the substrate, 
         whereas \emph{discrete} tethers constitute the connections in the models D--F. 
         In the models A and D a single membrane is attached at a substrate, 
         in B and E two membranes are linked together but are mobile in the vertical direction as a whole, 
         and in C and F a membrane pair is additionally attached to a substrate. 
         The single membrane models A and D have already been discussed in our previous work \cite{Merath_Seifert_PRE_2006} 
         and provide a basis for comparison of the different models in limiting cases. 
} 
\label{fig:1}
\end{figure*}

In a next step towards more biocompatible \cite{Boxer_CurrOpChemBiol_2000} soft interfaces and to more realistic biomembrane 
models one could attach a membrane not directly to a substrate but rather to a second, supported membrane. 
Since pre\-sently several experimental efforts are going into this direction 
\cite{Kaizuka_Groves_2004_and_2006,Parthasarathy_Groves_2004_and_2006,Leidy_Mouritsen_BiophysJ_2002}, 
we expect sup\-port\-ed membrane systems with mo\-re than one fluctuating membrane to be constructed in the near future 
(e.\,g., by utilizing vesicles tethered to a sup\-port\-ed membrane \cite{Yoshina-Ishii_Boxer_JACS_2003} 
and subsequent vesicle fusion). 
These sys\-tems could be used 
(i) to study diffusion within a membrane in a native-like environment, 
(ii) to tune, to a certain degree, the fluctuation properties of the con\-sti\-tu\-ent membranes 
in order to investigate fluc\-tu\-ation-de\-pen\-dent diffusion of membrane proteins 
\cite{Lin_Brown_BiophysJ_2004,Reister_Seifert_EPL_2005} 
or to suppress membrane fluctuations in order to avoid undesired contact \cite{Elie-Caille_Bourdillon_Langmuir_2005}, 
and 
(iii) to prolongate the lifetime of biosensors~\cite{Albertorio_Cremer_Langmuir_2005}.  
Contemporary passivations of the substrate usually last only a few days 
and the lower membrane could then act as an alternative passivation that prevents the upper membrane to contact the substrate.

The purpose of this article is twofold:
(i) The fluctuation behavior (in Fourier space and in real space) 
of a tethered two-membrane system on a substrate is analyzed. 
The discreteness of the tether sites demands numerical calculations even though the result can be expressed 
in a quasi-analytical way. 
(ii) Smearing out the discreteness of the tethers results in a \emph{continuous} confining potential. 
For this system the fluctuation spectrum for arbitrary ratios of membrane bending rigidities 
as well as the standard deviation in real-space for equal bending rigidities are displayed analytically. 
Typical membrane displacements as a function of the ratio of bending rigidities are shown. 
Continuously interacting membrane systems are interesting on its own, nethertheless we present it here as a limiting case 
to which the results of the discrete model are compared. 
The systems considered here are supposed to be far from unbinding transitions~\cite{Lipowsky_Leibler_1986,Netz_Lipowsky_PRL_1993}, 
thus steric repulsion can be subsumed in an effective harmonic binding potential~\cite{US_PRL_1995,Mecke_Charitat_Langmuir_2003}.

This article is organized as follows: 
In section \ref{sec:2}, we introduce the models addressed in our work.
In section \ref{sec:3}, the models containing continuous harmonic interactions
are considered, and fluctuation spectra are derived analytically. 
In section \ref{sec:4}, we approach systems comprising membranes connected by discrete tethers 
and derive quasi-analytic solutions for the corresponding fluctuation spectrum and root-mean-square displacement (RMSD). 
The results from the continuous and discrete model systems are compared. 
A summarizing perspective is given in section~\ref{sec:5}. 
Finally, the case of three membranes with continuous interactions is considered in the Appendix.

\section{Models}
\label{sec:2}
Several model systems of fluid compound membranes are analyzed in this article. 
In Fig.\ \ref{fig:1}, an overview of the different cases is given. 
The diversity of possible configurations is classified 
with respect to three structural features: We distinguish between 
(i) discrete tethers or continuous confinement, 
(ii) one or two (or even more) membranes 
and (iii) ``free'' (i.\,e., mobile as a whole) or supported systems. 

The behavior of a particular model can be understood qualitatively by considering limiting cases that correspond to 
other models from the above scheme. 
An interesting question is, under what circumstances the discrete nature of the tethers 
determines the fluctuation behavior significantly and makes numerical effort necessary. 
Thus, we compare the class of model systems with discrete tethers to corresponding systems with a 
continuous harmonic potential acting on elongation differences of the involved membranes.

Notations are explained exemplarily for model F:
The basal plane is given by a rectangle with side lengths $L_x$ and $L_y$, and  
the bending rigidities of the two membranes are denoted $\kappa_1$ and $\kappa_2$. 
The two equally large membrane surfaces are parametrized by 
$h^{\scriptscriptstyle (1)}(\vec{r}) \equiv h^{\scriptscriptstyle (1)}(x, y)$ (upper membrane)
and $h^{\scriptscriptstyle (2)}(\vec{r}) = h^{\scriptscriptstyle (2)}(x,y)$ (lower membrane), 
where $h^{\scriptscriptstyle (1)} = 0$ and $h^{\scriptscriptstyle (2)} = 0$ 
are the rest positions of the upper and lower membrane, respectively. 
In the rest position the two membranes are flat, parallel 
and separated by the rest length $l_0$ of the $N$ equally long, vertical tethers.  
Each tether between the two membranes is described by its lateral site $\vec{r}_\alpha$ and its spring constant $K_\alpha$. 
Each site of the $N^\mathrm{(s)}$ substrate--membrane tethers 
is given by $\vec{r}_\alpha^\mathrm{(s)}$ and its spring constant by $K_\alpha^\mathrm{(s)}$. 
In contrast to the discrete tethers, continuous harmonic interactions involve an elastic parameter $\gamma$ 
that can be related to a comparable spring constant (cf.\ equation~(\ref{eqn:gamma_K_Delta})).

In our theoretical model sytems the membranes can, in principle, penetrate each other and the substrate. 
Hence the results are only applicable to those systems for which such fluctuations are irrelevant. 
This is the case if the rest length $l_0$ of the tethers is considerably larger than typical membrane elongations 
which can be estimated a posteriori by the root-mean-square displacement.

\section{Membranes with continuous interacting potential}
\label{sec:3}
We first study systems with interactions between a membrane and a nearly parallel object 
like another membrane, a substrate or an underlying cytoskeleton. 
Here, these interactions will be approximated by a continuous, harmonic, confining potential.  
Such models with continuous interactions also serve as comparison 
with the discrete tether-linked models addressed in the next section.

\subsection{Model B: Membrane pair with continuous interaction}
\label{subsec:2M_conti} 
As the simplest two-membrane model system the case of two membranes connected by a continuous binding potential
(model B, cf.\ Fig.~\ref{fig:1}) is studied. 
The Hamiltonian of the system consists of a Helfrich bending energy term for each membrane 
and the potential energy due to the interaction, 
\begin{eqnarray}   \label{eqn:Def_H(M2,conti)}
   H^{\scriptscriptstyle \mathrm{(B)}}  & \equiv &  
       \int_{0}^{L_x}  \!\!\!\!\! \d x   \int_{0}^{L_y}  \!\!\!\!\! \d y 
                 \Big\{      \frac{\kappa_1}{2}  \big[\nabla^2 h^{\scriptscriptstyle (1)}(\vec{r}) \big]^2    
     +    \frac{\kappa_2}{2} 
                 \big[\nabla^2 h^{\scriptscriptstyle (2)}(\vec{r}) \big]^2     \nonumber \\
 &&   + {} \frac{\gamma}{2}  \big[ h^{\scriptscriptstyle (1)}(\vec{r}) 
                       - h^{\scriptscriptstyle (2)}(\vec{r})  \big]^2  \Big\}  \ \ .
\end{eqnarray}
The index $1$ or $2$ refers to the upper or lower membrane, respectively. 
The parameter $\gamma$ determines the strength of the harmonic potential. 
A Fourier ansatz for both heigth profiles, 
\begin{equation}  \label{eqn:Fourier_ansatz}
   h^{\scriptscriptstyle (i)}(\vec{r}) \: = \: \sum\nolimits_{\vec{k}} h_\vec{k}^{\scriptscriptstyle (i)} 
                                           \, e^{i \vec{k}\cdot\vec{r}} \, ,
\end{equation}
with $i \in \{1, 2\}$, $\vec{k}  \equiv  (k_x, k_y)$ and $k_{x,y}  = (0, \pm 1, \pm 2, \ldots) \cdot (2 \pi / L_{x,y})$, 
leads to 
\begin{eqnarray} 
   H^{\scriptscriptstyle \mathrm{(B)}}  & = &  
       \frac{1}{2} L_x L_y \sum\nolimits_{\vec{k}} \Big[
                ( \kappa_1 |\vec{k}|^4 + \gamma ) |h_\vec{k}^{\scriptscriptstyle (1)}|^2          \nonumber \\
 &&    {}  + ( \kappa_2 |\vec{k}|^4 + \gamma ) |h_\vec{k}^{\scriptscriptstyle (2)}|^2
           - 2 \gamma \, \mathrm{Re}\big( h_\vec{k}^{\scriptscriptstyle (1)} \, {h_\vec{k}^{\scriptscriptstyle (2)}}^* \big)  \Big]
                \, , ~
\end{eqnarray}
where $\mathrm{Re}$ denotes the real part. 
Separating the Fourier coefficients into real and imaginary parts, 
\begin{equation}  \label{eqn:real_and_imaginary}
   h_\vec{k}^{\scriptscriptstyle (i)} 
    \: \equiv \:     a_\vec{k}^{\scriptscriptstyle (i)} + i \, b_\vec{k}^{\scriptscriptstyle (i)   }\quad 
                    \mathrm{for~} i\in\{1,2\} \, ,
\end{equation}
results in 
\begin{eqnarray} 
   H^{\scriptscriptstyle \mathrm{(B)}}  & = &  
      \frac{1}{2} L_x L_y \sum\nolimits_{\vec{k}} \Big[
         ( \kappa_1 |\vec{k}|^4 + \gamma ) \Big( {a_\vec{k}^{\scriptscriptstyle (1)}}^2  + 
                 {b_\vec{k}^{\scriptscriptstyle (1)}}^2 \Big)
                            \nonumber \\
 &&   {}  + ( \kappa_2 |\vec{k}|^4 + \gamma ) \Big( {a_\vec{k}^{\scriptscriptstyle (2)}}^2  + 
                 {b_\vec{k}^{\scriptscriptstyle (2)}}^2 \Big)
                            \nonumber \\
 &&   {}  - 2 \gamma  \Big( a_\vec{k}^{\scriptscriptstyle (1)} a_\vec{k}^{\scriptscriptstyle (2)} 
                             + b_\vec{k}^{\scriptscriptstyle (1)}  b_\vec{k}^{\scriptscriptstyle (2)} \Big)  \Big]  \ \ .
\end{eqnarray}
Since the height profiles are real, the Fourier coefficients fulfill the conditions 
$a_\vec{-k}^{\scriptscriptstyle (i)} = a_\vec{k}^{\scriptscriptstyle (i)}$ and 
$b_\vec{-k}^{\scriptscriptstyle (i)} = - b_\vec{k}^{\scriptscriptstyle (i)}$, 
so that we chose the coefficients belonging to the wave vectors $\{\vec{q}\}$ and $\vec{0}$ as independent variables. 
In order to avoid divergent translational modes, we exclude the translational mode $a_\vec{0}^{\scriptscriptstyle (2)}$ of the 
lower membrane. 
Or, in other words, the vertical position $a_\vec{0}^{\scriptscriptstyle (2)}$ serves as frame of reference. 
This choice is motivated by related model systems containing a substrate: 
Since a substrate on the one hand calls for acting as frame of reference 
and on the other hand corresponds to an infinitely rigid membrane, 
we do not turn to the center of mass as frame of reference but to the spatial average of the lower membrane height profile. 
Thus the Hamiltonian as a functional of the independent variables is 
\begin{eqnarray}  \label{eqn:independent_2co}
   H^{\scriptscriptstyle \mathrm{(B)}}  & = &  
      L_x L_y \sum\nolimits_{\vec{q}} \Big[
         ( \kappa_1 |\vec{q}|^4 + \gamma ) \Big( {a_\vec{q}^{\scriptscriptstyle (1)}}^2  
                     + {b_\vec{q}^{\scriptscriptstyle (1)}}^2 \Big)
                            \nonumber \\
 &&   {}  + ( \kappa_2 |\vec{q}|^4 + \gamma ) \Big( {a_\vec{q}^{\scriptscriptstyle (2)}}^2  
                     + {b_\vec{q}^{\scriptscriptstyle (2)}}^2 \Big)
                            \nonumber \\
 &&   {}  - 2 \gamma  \big( a_\vec{q}^{\scriptscriptstyle (1)} a_\vec{q}^{\scriptscriptstyle (2)} 
                        + b_\vec{q}^{\scriptscriptstyle (1)}  b_\vec{q}^{\scriptscriptstyle (2)} \big)  \Big]
     + \frac{1}{2} L_x L_y \gamma {a_\vec{0}^{\scriptscriptstyle (1)}}^2     {} .
\end{eqnarray}

Integrating out the degrees of freedom of the lower membrane,
$\{a_\vec{q}^{\scriptscriptstyle (2)}\}$ and $\{b_\vec{q}^{\scriptscriptstyle (2)}\}$, 
leads to an effective Hamiltonian for the upper membrane, 
\begin{eqnarray} 
   e^{-\beta H_\mathrm{eff}^{\scriptscriptstyle (1)}} 
      & \equiv &  \int {\mathcal D}h^{\scriptscriptstyle (2)} e^{- \beta H^{\scriptscriptstyle \mathrm{(B)}}}  \\
      & = & \bigg( \prod\nolimits_{\vec{q}} 
          \int_{- \infty}^{\infty}  \!\!\!\!\! \d a_\vec{q}^{\scriptscriptstyle (2)}  
          \int_{- \infty}^{\infty}  \!\!\!\!\! \d b_\vec{q}^{\scriptscriptstyle (2)}  \bigg) 
          e^{- \beta H^{\scriptscriptstyle \mathrm{(B)}}}    \ .   \label{eqn:Def_H_eff_1}
\end{eqnarray}
Applying the auxiliary formula 
\begin{eqnarray} 
 &&  \int_{- \infty}^{\infty}  \!\!\!\!\! \d z^{\scriptscriptstyle (2)}  
     e^{- \beta L_x L_y ( \kappa_2 |\vec{q}|^4 + \gamma ) \, {z^{\scriptscriptstyle (2)}}^2 
         - \beta L_x L_y (-2 \gamma) \,   z^{\scriptscriptstyle (1)}   z^{\scriptscriptstyle (2)} }  \nonumber  \\
 &   \, \sim \,   &   \exp\left( \frac{\beta L_x L_y }{ \kappa_2 |\vec{q}|^4 + \gamma } \gamma^2 \, 
               {z^{\scriptscriptstyle (1)}}^2  \right)
\end{eqnarray} 
for each integral in equation (\ref{eqn:Def_H_eff_1}) 
gives 
\begin{eqnarray} 
  &&  e^{-\beta H_\mathrm{eff}^{\scriptscriptstyle (1)}}       \nonumber \\
  & \sim &  \prod\nolimits_{\vec{q}}  e^{ - \beta L_x L_y \big(\kappa_1 |\vec{q}|^4 + \gamma   
                                          - \frac{\gamma^2}{\kappa_2 |\vec{q}|^4 + \gamma} \big) 
              \big( {a_\vec{q}^{\scriptscriptstyle (1)}}^2  + {b_\vec{q}^{\scriptscriptstyle (1)}}^2 \big)  } \  .
\end{eqnarray} 
After a suitable shift of the energy zero-level, the effective Hamiltonian reads 
\begin{eqnarray} 
  H_\mathrm{eff}^{\scriptscriptstyle (1)}    
  & = &  L_x L_y \Big( \kappa_1 |\vec{q}|^4 + \gamma   - \frac{\gamma^2}{\kappa_2 |\vec{q}|^4 + \gamma} \Big) \,   \nonumber \\
  &&        \times     \big( {a_\vec{q}^{\scriptscriptstyle (1)}}^2  + {b_\vec{q}^{\scriptscriptstyle (1)}}^2 \big)  
     \; + \frac{1}{2} L_x L_y \gamma \, {a_\vec{0}^{\scriptscriptstyle (1)}}^2   \  .
\end{eqnarray}

The fluctuation spectrum of the upper membrane can now be read off. 
For non-vanishing wave vectors it is given by 
\begin{eqnarray}  \label{eqn:spectrum_2co_not0}
   \langle \, |h^{\scriptscriptstyle (1)}_{\vec{k} \neq \vec{0}}|^2 \, \rangle & = & 
             \langle \, {a^{\scriptscriptstyle (1)}_{\vec{k}}}^2 \, \rangle  
       +     \langle \, {b^{\scriptscriptstyle (1)}_{\vec{k}}}^2 \, \rangle
  \; = 2 \;   \langle \, {a^{\scriptscriptstyle (1)}_{\vec{k}}}^2 \, \rangle   \nonumber  \\
   & = &  \frac{k_B T / L_x L_y}{ \kappa_1 |\vec{k}|^4 + \gamma   - \frac{\textstyle \gamma^2}{\textstyle \kappa_2 |\vec{k}|^4 
                   + \gamma}}   \\
   & = &  \frac{k_B T / L_x L_y}{ \kappa_1 k^4 +  \gamma_{\mathrm{eff}}(k) }  \  ,
\end{eqnarray} 
with an effective potential parameter 
\begin{equation}  \label{eqn:Def_gamma_eff}
   \frac{1}{ \gamma_{\mathrm{eff}}(k) }  ~ \equiv ~ \frac{1}{\gamma}  \,+\,  \frac{1}{\kappa_2 \, k^4}  
\end{equation}
and the abbreviation $k \equiv |\vec{k}|$. 
This formula does not apply for the translational mode of the upper membrane, i.\,e., for the spectral origin $\vec{k} = \vec{0}$,
since $a_\vec{0}^{\scriptscriptstyle (2)}$ was set to zero manually. Instead, at the origin the spectrum attains the value 
\begin{eqnarray} 
   \langle \, |h^{\scriptscriptstyle (1)}_{\vec{0}}|^2 \, \rangle & = & 
             \langle \, {a^{\scriptscriptstyle (1)}_{\vec{0}}}^2 \, \rangle  
   \: = \:     \frac{k_B T / L_x L_y}{ \gamma }  \ \, .
\end{eqnarray} 

If the translational mode $a_\vec{0}^{\scriptscriptstyle (2)}$ of the lower membrane was permitted 
(i.\,e., if it was regarded as a degree of freedom), 
the spectrum would be described correctly for all wave vectors by 
\begin{eqnarray} 
   \langle \, |h^{\scriptscriptstyle (1)}_{\vec{k}}|^2 \, \rangle 
   & = &  \frac{k_B T / L_x L_y}{ \kappa_1 \, k^4 +  \gamma_{\mathrm{eff}}(k) }  \  ,
\end{eqnarray} 
featuring a divergence at the origin.

\begin{figure}[!tb]
\hfil \hspace{-0.0cm}  \includegraphics[scale=0.55, clip]{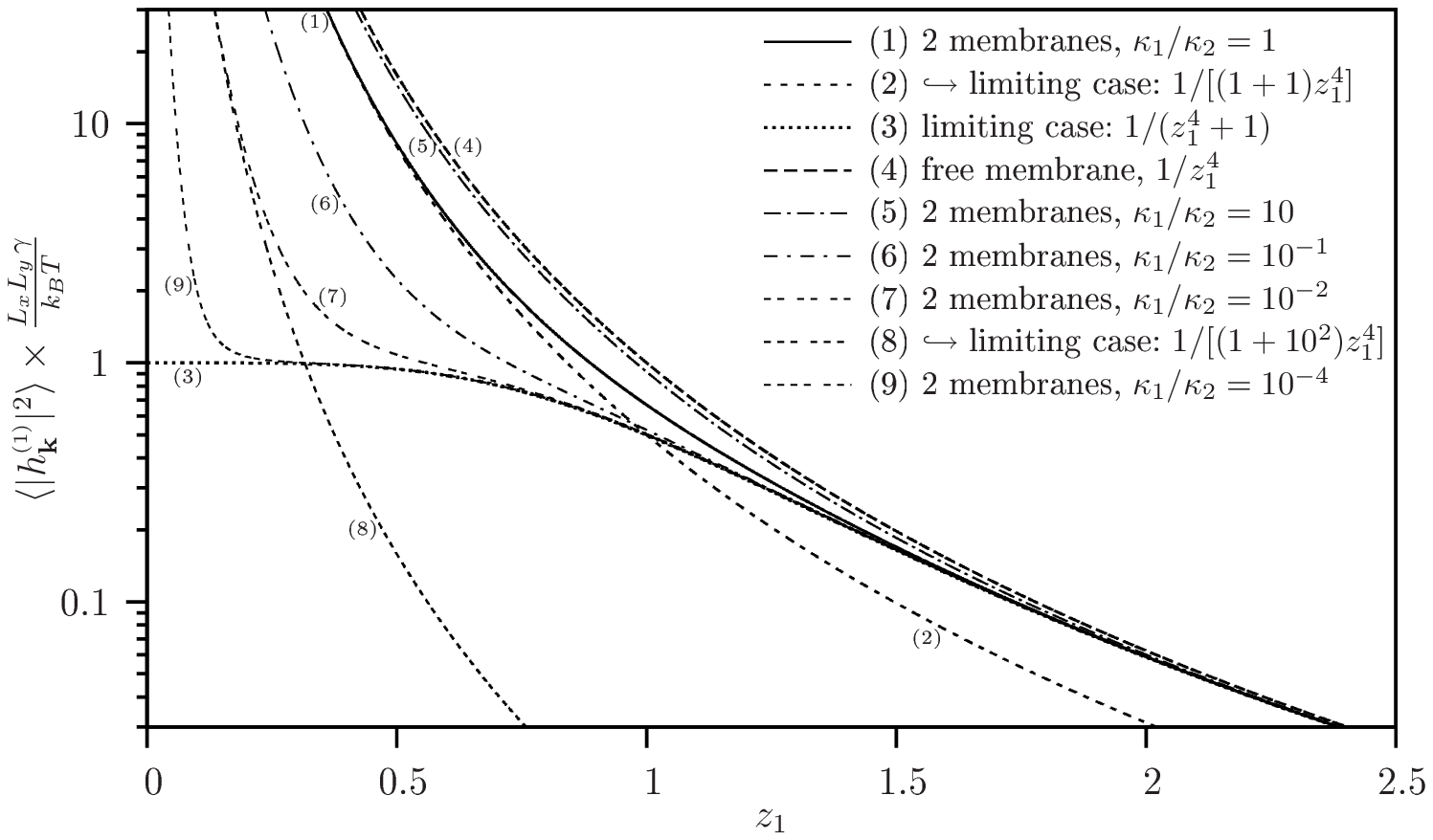}   \hfil
\caption{Fluctuation spectra for model B. 
         A membrane is continuously, harmonically coupled to an almost parallel second (lower) membrane. 
         For several values of the ratio $\kappa_1/\kappa_2$ of the bending rigidities, 
         the scaled spectrum is plotted versus the reduced wave vector $z_1$ 
         containing the bending rigidity $\kappa_1$ of the upper membrane and the stiffness parameter $\gamma$ 
         of the harmonic potential. 
         At the origin the scaled spectrum always diverges, if the translational mode of the lower membrane is permitted, 
         whereas it equals 1, if this mode is artificially excluded 
         or the lower membrane is regarded as the center of the frame of reference. 
         The limiting curves for large or small $z_1$, respectively, intersect at $z_1 = (\kappa_1/\kappa_2)^{1/4}$, 
         indicating the position of the cross-over region. 
         Evidently, the spectrum becomes similar to the spectrum of a single membrane attached to a substrate 
         for $\kappa_2 \to \infty$,
         and the spectrum of a free membrane is recovered for $\kappa_2 \to 0$. 
} 
\label{fig:2M_conti}
\end{figure}

Examining $\gamma_\mathrm{eff}$ from equation (\ref{eqn:Def_gamma_eff}) a striking analogy is emerging. 
$\gamma_\mathrm{eff}$ depends on $\gamma$ and $\kappa_2 k^4$ 
like the effective spring constant of two springs in series depends on the constituent spring constants. 
Thus, the lower membrane affects the upper membrane like a second continuous spring, 
arranged in series to the connecting continuous springs. 
In contrast to the ``real'' continuous spring parameter $\gamma$, 
the strength $\kappa_2 k^4$ of the ``virtual'' continuous springs depends on the wavelength. 

The limiting cases for $\gamma_\mathrm{eff}$ are 
\begin{eqnarray} 
   \gamma_\mathrm{eff}  & \simeq &  \left\{  \begin{array}[c]{cl}   \gamma 
                                         & \quad \mathrm{for~} \gamma \ll  \kappa_2 \, k^4  \\[0.9ex] 
                         \kappa_2 \, k^4  
                                         & \quad \mathrm{for~} \gamma \gg  \kappa_2 \, k^4   \  ,  \end{array}  \right.
\end{eqnarray} 
and thus for the spectrum 
\begin{eqnarray} \label{eqn:limits_spectrum}
   \langle \, |h^{\scriptscriptstyle (1)}_{\vec{k}}|^2 \, \rangle 
    & \simeq &  \left\{  \begin{array}[c]{cl}   \displaystyle
           \frac{k_B T / L_x L_y}{ \kappa_1 \, k^4 +  \gamma }                & 
                                   \quad \mathrm{for~}    k \gg  (\gamma / \kappa_2)^{1/4} \\[2.4ex] 
           \frac{k_B T / L_x L_y}{ (\kappa_1 + \kappa_2) \, k^4 }  & 
                                   \quad \mathrm{for~}    k \ll  (\gamma / \kappa_2)^{1/4} \ \, .  \end{array}  \right.  
\end{eqnarray} 
Consequently, for large wave vectors, the case of a single membrane linked to a substrate by continuous springs is recovered, 
while for small wave vectors, the upper membrane behaves as if two free membranes were glued together 
exhibiting an effective bending rigidity 
$\kappa_1 + \kappa_2$.

Introducing the reduced wave vector 
\begin{eqnarray}  \label{eqn:Def_reduced_wavevector}
   z_1  & \equiv &  k \, \xi_1   \ ,
\end{eqnarray}
with the persistence length 
\begin{eqnarray}
   \xi_1 & \equiv & (\kappa_1 / \gamma)^{1/4}   \ ,
\end{eqnarray}
the spectrum can be expressed as 
\begin{eqnarray} 
   \langle \, |h^{\scriptscriptstyle (1)}_{\vec{k}}|^2 \, \rangle 
   & = &  \frac{k_B T}{\gamma \, L_x L_y}   \times  \frac{1}{ z_1^4  +  1  -  \frac{\textstyle 1}{ \textstyle \frac{\textstyle
   \kappa_2}{\textstyle \kappa_1} z_1^4 + 1 }  }    \ .
\end{eqnarray} 
Its limiting cases are 
\begin{eqnarray} 
  \frac{ \langle \, |h^{\scriptscriptstyle (1)}_{\vec{k}}|^2 \, \rangle }{ \frac{k_B T}{\gamma \, L_x L_y} }
    & \simeq &  \left\{  \begin{array}[c]{ll}  \displaystyle
           \frac{1}{z_1^4  +  1 }   & 
                   \quad \mathrm{for~}   z_1  \gg  \big(\frac{\textstyle \kappa_1}{\textstyle \kappa_2}\big)^{1/4}  \\[1.8ex] 
           \displaystyle
           \frac{1}{ (1 + \frac{\kappa_2}{\kappa_1}) z_1^4  }   & 
                      \quad \mathrm{for~}   z_1  \ll  \big(\frac{\textstyle \kappa_1}{\textstyle \kappa_2}\big)^{1/4}   \ . 
                          \end{array}  \right.  
\end{eqnarray} 
The spectrum of the upper membrane and its limiting cases are displayed in Fig.\ \ref{fig:2M_conti} 
for varying values of $\kappa_1/\kappa_2$.

\subsection{Model C: Membrane pair on a substrate, with continuous interactions} 
\label{subsec:2M_substrate_conti} 
If the membrane pair from the previous subsection (model B) is attached to a substrate by a continuous harmonic interaction,  
the Hamil\-ton\-ian reads 
\begin{eqnarray}  \label{eqn:Def_H2co_s}
   H^{\scriptscriptstyle \mathrm{(C)}}  & \equiv &  
       \int_{0}^{L_x}  \!\!\!\!\! \d x   \int_{0}^{L_y}  \!\!\!\!\! \d y 
                 \Big\{      \frac{\kappa_1}{2}  \big[\nabla^2 h^{\scriptscriptstyle (1)}(\vec{r}) \big]^2    
     +    \frac{\kappa_2}{2} 
                 \big[\nabla^2 h^{\scriptscriptstyle (2)}(\vec{r}) \big]^2     \nonumber \\
 &&  +    \frac{\gamma_1}{2}  \big[ h^{\scriptscriptstyle (1)}(\vec{r}) 
                       - h^{\scriptscriptstyle (2)}(\vec{r})  \big]^2 
     +    \frac{\gamma_2}{2}  \big[ h^{\scriptscriptstyle (2)}(\vec{r}) \big]^2   \Big\}   \ \ .
\end{eqnarray}
This situation is equivalent to a system of three membranes with continuous interacting potentials 
(cf.\ the App\-en\-dix), 
in which the bending rigidity of the lowest membrane is set to infinity, $\kappa_3 \to \infty$, 
and the translation mode $a_\vec{0}^{\scriptscriptstyle (3)}$ is excluded. 
The fluctuation spectra of the upper membrane, 
\begin{eqnarray} 
   \langle \, |h^{\scriptscriptstyle (1)}_\vec{k}|^2 \, \rangle 
   & = &  \frac{k_B T / L_x L_y}{ \kappa_1 \, k^4 + \gamma_1   
                  - \frac{\textstyle \gamma_1^2}{\textstyle \kappa_2 \, k^4 + \gamma_1 + \gamma_2   } }   \ ,
\end{eqnarray} 
and of the lower membrane, 
\begin{eqnarray}
   \langle \, |h^{\scriptscriptstyle (2)}_\vec{k}|^2 \, \rangle 
   & = &  \frac{k_B T / L_x L_y}{ \kappa_2 \, k^4          + \gamma_2
                  + \frac{\textstyle 1}{\textstyle   \frac{\textstyle 1}{\textstyle \kappa_1 \, k^4} 
                                                   + \frac{\textstyle  1}{\textstyle \gamma_1} } }      \ ,
\end{eqnarray} 
follow from equations (\ref{eqn:3M_co_top}) and (\ref{eqn:3M_co_middle}), respectively, 
and also hold at the origin $\vec{k} = \vec{0}$.

\begin{figure}[!tb]
\hfil \hspace{-0.0cm}  \includegraphics[scale=0.62, clip]{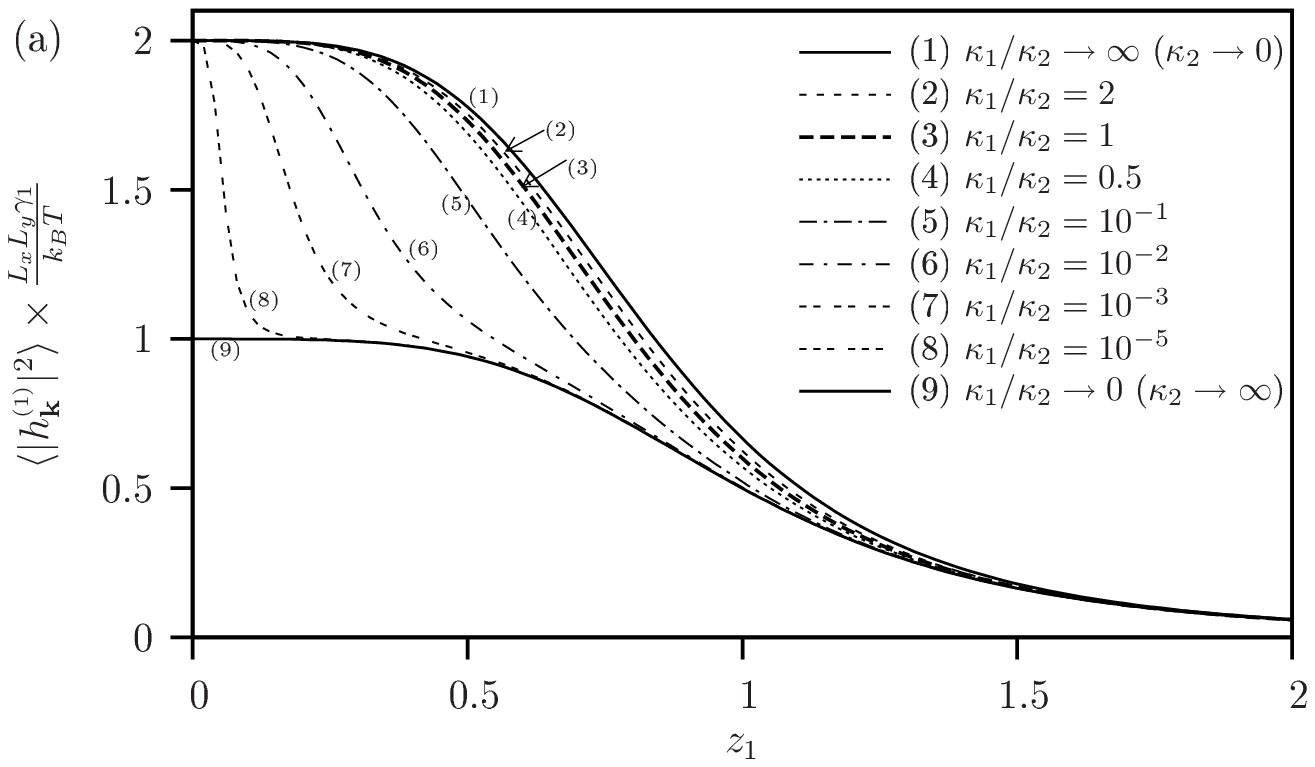}   \hfil

\hfil \hspace{-0.0cm}  \includegraphics[scale=0.62, clip]{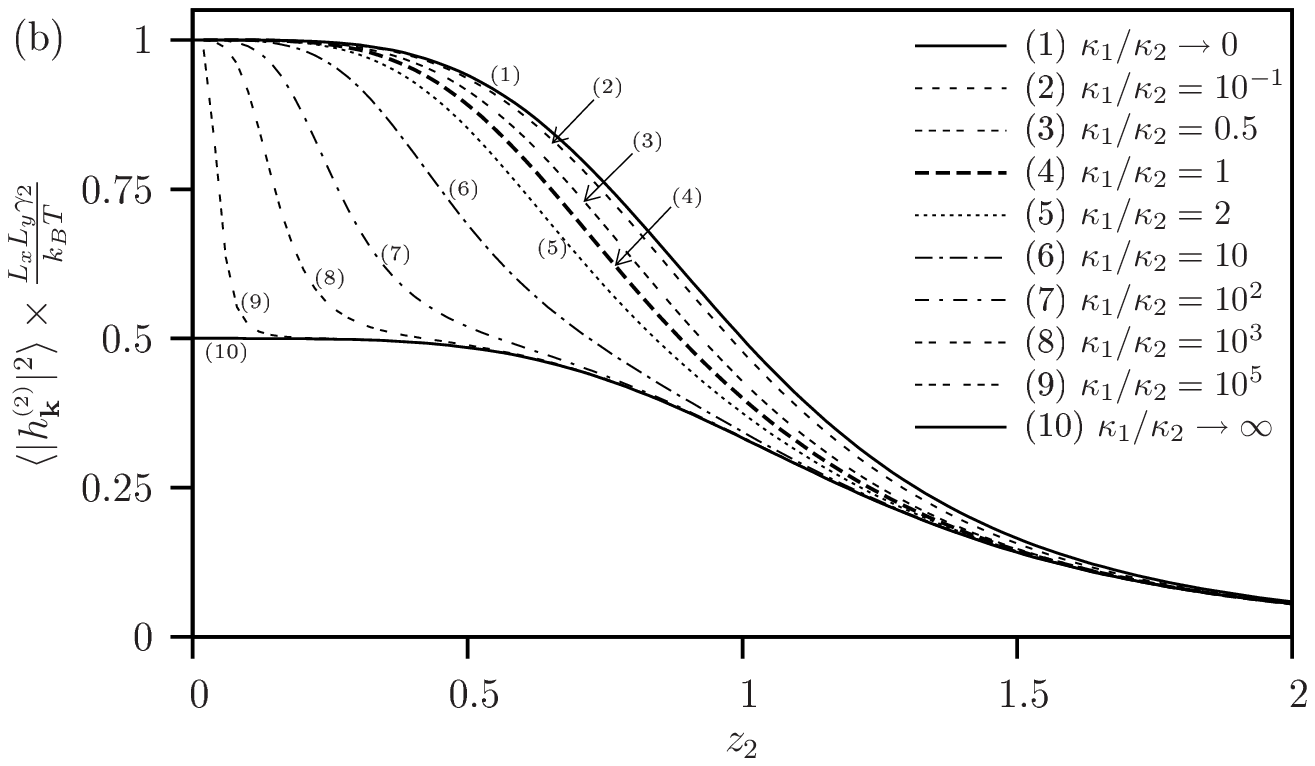}   \hfil
\caption{Fluctuation spectra of the upper (a) and lower (b) membrane in model C, i.\,e., in a system consisting of two membranes 
         coupled among each other and to a substrate via a continuous interacting potential. 
         The scaled spectra (cf.\ equations (\ref{eqn:scaled_spectrum_top}) and (\ref{eqn:scaled_spectrum_bottom})) 
         are plotted versus the reduced wave vector $z_1$ and $z_2$ (cf.\ equation (\ref{eqn:z_von_k})), respectively. 
         Both elastic constants are set equal, $\gamma_1 = \gamma_2$.   \newline
         (a) The spectrum of the upper membrane recovers the spectrum of a single membrane attached to a substrate 
             with $\gamma = 1/2 \, \gamma_1$ for $\kappa_2 \to 0$ and with $\gamma = \gamma_1$ for $\kappa_2 \to \infty$ 
             (except at the origin, since the translational mode $h^{\scriptscriptstyle (1)}_\vec{0}$ 
             is not affected by the bending rigidity $\kappa_2$ 
             and obeys $\langle \, |h^{\scriptscriptstyle (1)}_\vec{0}|^2 \,\rangle = 2 \times k_B T/L_x L_y  \gamma_1$). ---
         (b) The lower membrane complies with a single membrane at a substrate with $\gamma = \gamma_1$ for $\kappa_1 \to 0$ 
             and with $\gamma = 2 \, \gamma_1$ for $\kappa_1 \to \infty$ 
             (except at the origin, since $\kappa_1$ does not hinder the translational mode from maintaining 
             $\langle \, |h^{\scriptscriptstyle (2)}_\vec{0}|^2 \,\rangle = 1 \times k_B T/L_x L_y  \gamma_2$). 
} 
\label{fig:2M_substrate_conti}
\end{figure}

The spectra above can be written as 
\begin{eqnarray}  \label{eqn:scaled_spectrum_top}
   \langle \, |h^{\scriptscriptstyle (1)}_\vec{k}|^2 \, \rangle 
   & = &  \frac{k_B T}{L_x L_y \, \gamma_1}  \times \frac{1}{ z_1^4 + 1 
                  - \frac{\textstyle 1}{\textstyle  \frac{\textstyle \kappa_2}{\textstyle \kappa_1} z_1^4 + 1 + \frac{\textstyle \gamma_2}{\textstyle \gamma_1}   } }   
\end{eqnarray} 
and 
\begin{eqnarray}  \label{eqn:scaled_spectrum_bottom}
   \langle \, |h^{\scriptscriptstyle (2)}_\vec{k}|^2 \, \rangle 
   & = &  \frac{k_B T}{L_x L_y \, \gamma_2}  \times \frac{1}{ z_2^4    + 1 
                  + \frac{\textstyle 1}{\textstyle   \frac{\textstyle 1}{\textstyle  \frac{\textstyle \kappa_1}{\textstyle \kappa_2} z_2^4} + \frac{\textstyle 1}{\textstyle  \frac{\textstyle \gamma_1}{\textstyle \gamma_2} } } }      \ ,
\end{eqnarray} 
with the reduced wave vectors ($i \in \{1, 2\}$) 
\begin{eqnarray}  \label{eqn:z_von_k}
   z_i  & \equiv & k \, \xi_i 
\end{eqnarray}
and the persistence lengths 
\begin{eqnarray}
   \xi_i & \equiv & (\kappa_i / \gamma_i)^{1/4}   \ .
\end{eqnarray}
On increasing $\gamma_i$, the spectrum $\langle \, |h^{\scriptscriptstyle (i)}_\vec{k}|^2 \, \rangle$ 
is stretched towards higher wave vectors and its amplitude is reduced. 
In contrast to the tether models discussed below, the continuous harmonic potential does not involve any physical length, 
and the length scale of the system is merely given by the persistence lengths $\xi_i$.
For the simplified case of equally strong interacting potentials, 
i.\,e., $\gamma \equiv \gamma_1 = \gamma_2$, 
these two types of spectra are displayed for various values of $\kappa_1/\kappa_2$ 
in Fig.~\ref{fig:2M_substrate_conti}.

The \emph{real-space fluctuations} of this system are con\-si\-der\-ed now. 
In general, the mean-square dis\-place\-ment (MSD) of the membrane $i$ 
results from the correla\-tions of the Fourier coefficients, 
\begin{eqnarray}  \label{eqn:MSD_general}
   \langle \, \big[ h^{\scriptscriptstyle (i)}(\vec{r}) \big]^2 \, \rangle 
   & = &    \sum_{\vec{k}, \vec{k}^\prime}   \langle \, 
            h^{\scriptscriptstyle (i)}_\vec{k} h^{\scriptscriptstyle (i)}_{\vec{k}^\prime}    \, \rangle 
            \, e^{i (\vec{k} + \vec{k}^\prime)\cdot \vec{r} }  \ .
\end{eqnarray} 
Since the present system with continuous interacting potential is translationally invariant, it is 
\begin{equation}
\langle \, h^{\scriptscriptstyle (i)}_\vec{k} h^{\scriptscriptstyle (i)}_{\vec{k}^\prime}    \, \rangle 
=  \langle \, |h^{\scriptscriptstyle (i)}_\vec{k}|^2 \, \rangle  \, \delta^{~}_{\vec{k}^\prime,-\vec{k}}    \ .
\end{equation}
Consequently, the MSD of the upper ($i=1$) and lower ($i=2$) membrane can be expressed as 
\begin{eqnarray} 
   \langle \, \big[ h^{\scriptscriptstyle (i)}(\vec{r}) \big]^2 \, \rangle 
   & = &    \sum_\vec{k}   \langle \, |h^{\scriptscriptstyle (i)}_\vec{k}|^2   \, \rangle   \nonumber \\
   & \to &  \frac{L_x\,L_y}{4 \pi^2}   \int\nolimits_{\mathbb{R}^2}  \d \vec{k} 
               \langle \, |h^{\scriptscriptstyle (i)}_\vec{k}|^2   \, \rangle   \ .
\end{eqnarray}

\begin{figure}[!tb]
\hfil \hspace{-0.1cm}  \includegraphics[scale=0.65, clip]{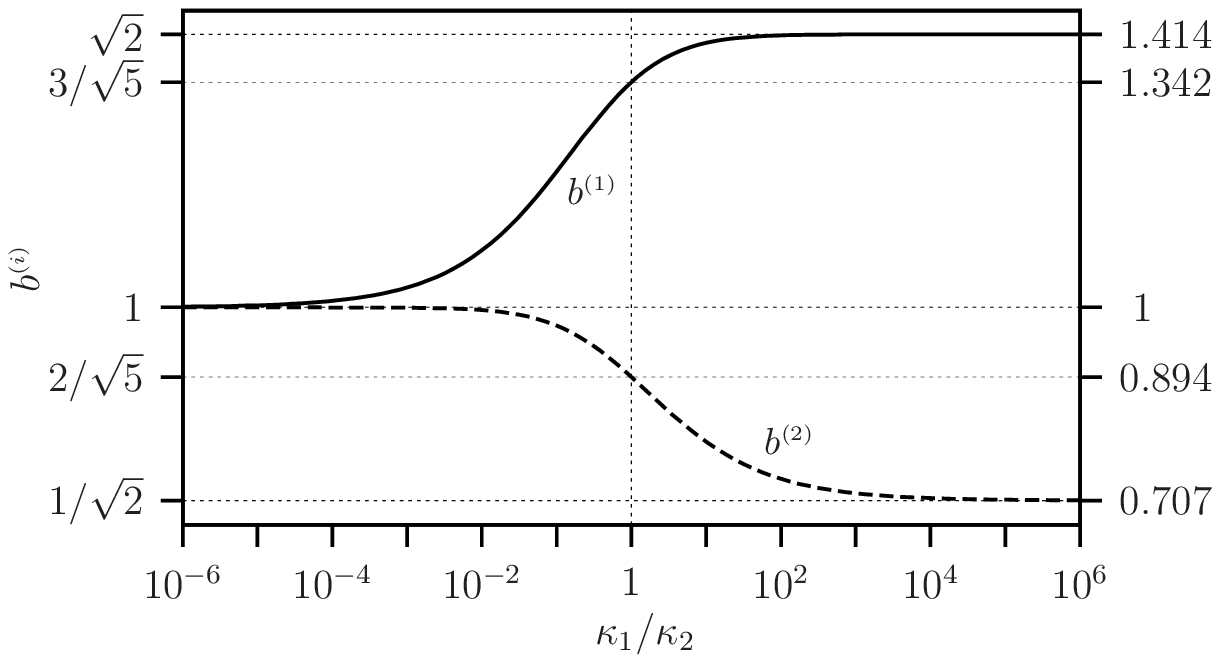}   \hfil
\caption{Scaling functions $b^{\scriptscriptstyle (1)}( \kappa_1 / \kappa_2 )$ 
         and $b^{\scriptscriptstyle (2)}( \kappa_1 / \kappa_2 )$ 
         of the upper or lower membrane, respectively, in model C. 
         They describe the amplification of the MSD in a system of two membranes with a continuous binding potential 
         among each other and towards a substrate with respect to a single membrane attached to a substrate. 
         Here, the two involved continuous potentials are assumed to be equally strong. 
} 
\label{fig:2M_substrate_conti_MSD}
\end{figure}

We consider the special case of equally strong interacting potentials, i.\,e., 
$\gamma \equiv \gamma_1 = \gamma_2$. 
After simple manipulations, the MSD of the upper membrane reads 
\begin{eqnarray}   \label{eqn:MSD_2Mc_1}
   \langle \, \big[ h^{\scriptscriptstyle (1)}(\vec{r}) \big]^2 \, \rangle 
   & = &  \frac{k_B T}{8 \sqrt{\gamma \kappa_1}}  \times   \frac{2}{\pi}  \int\nolimits_{0}^{\infty} \!\!\!\! \d x 
          \frac{x^2 + 2 \frac{\kappa_1}{\kappa_2}}{x^4 + (1+2 \frac{\kappa_1}{\kappa_2}) x^2 
             + \frac{\kappa_1}{\kappa_2}}   \nonumber \\ 
   & \equiv & \frac{k_B T}{8 \sqrt{\gamma \kappa_1}}  \times  b^{\scriptscriptstyle (1)}( \kappa_1 / \kappa_2 )  \ ,
\end{eqnarray} 
where the first factor corresponds to the MSD of one single membrane linked to a substrate 
by an interacting potential with parameter $\gamma$ 
and the scaling function $b^{\scriptscriptstyle (1)}$ describes the influence of the lower on the upper membrane. 
One can read off the two limiting cases, 
\begin{eqnarray}  \label{eqn:limitingcases1}
   b^{\scriptscriptstyle (1)}( \kappa_1 / \kappa_2 )
    & \to &  \left\{  \begin{array}[c]{cl}   1       &   \quad \mathrm{for~}    \kappa_2 \to \infty   \\
                                           \sqrt{2}  &   \quad \mathrm{for~}    \kappa_2 \to 0    \ .   \end{array}  \right.  
\end{eqnarray} 
Analogously, the MSD of the lower membrane reads 
\begin{eqnarray}  \label{eqn:MSD_2Mc_2}
   \langle \, \big[ h^{\scriptscriptstyle (2)}(\vec{r}) \big])^2 \, \rangle 
   & = &  \frac{k_B T}{8 \sqrt{\gamma \kappa_2}}  \times   \frac{2}{\pi}  \int\nolimits_{0}^{\infty} \!\!\!\! \d x 
          \frac{x^2 + \frac{\kappa_2}{\kappa_1}}{x^4 + (2 + \frac{\kappa_2}{\kappa_1}) x^2 
             + \frac{\kappa_2}{\kappa_1}}   \nonumber \\ 
   & \equiv & \frac{k_B T}{8 \sqrt{\gamma \kappa_2}}  \times  b^{\scriptscriptstyle (2)}( \kappa_1 / \kappa_2 )  \ .
\end{eqnarray} 
The two limiting cases of the scaling function $b^{\scriptscriptstyle (2)}$ are 
\begin{eqnarray} \label{eqn:limitingcases2}
   b^{\scriptscriptstyle (2)}( \kappa_1 / \kappa_2 )
    & \to &  \left\{  \begin{array}[c]{cl}   1       &   \quad \mathrm{for~}    \kappa_1 \to 0   \\ 
                                           1/\sqrt{2}  &   \quad \mathrm{for~}    \kappa_1 \to \infty    \ .   \end{array}  \right.  
\end{eqnarray} 
For the case of equally rigid membranes, $\kappa_1 = \kappa_2$, the scaling functions take the values 
$b^{\scriptscriptstyle (1)}( 1 ) = 3/\sqrt{5} \simeq 1.342$ 
and $b^{\scriptscriptstyle (2)}( 1 ) = 2/\sqrt{5} \simeq 0.894$. 
The scaling functions $b^{\scriptscriptstyle (1)}$ and $b^{\scriptscriptstyle (1)}$ 
are shown in Fig.\ \ref{fig:2M_substrate_conti_MSD}. 
Compared to the single-membrane case, the fluctuations of the upper membrane are increased and those of the lower membrane are decreased. 
This result has an impact on related experimental systems, 
since the risk of undesired adsorption of the membrane at the substrate is reduced by introducing another membrane on top.

\section{Two membranes linked by (discrete) tethers}
\label{sec:4}
After having analyzed the models in which the membranes are subject to a continuous harmonic confinement, 
we address the models containing localized connections via \mbox{tethers}. 
First in subsection \ref{subsec:model_A1}, the basic formulae are derived for model E (cf.\ Fig.~\ref{fig:1}), 
where a coupled membrane pair can move freely in the vertical direction. 
Subsequently a solid support, composed of tethers as well, is added (model F) 
and incorporated into the calculation scheme (subsection \ref{subsec:solid_support}). 
Then the results are discussed with respect to fluctuation spectra (subsection \ref{subsec:spectra_discrete}) 
and typical membrane elongations (subsection~\ref{subsec:MSD_discrete}).

\subsection{Model E: Basic calculation}
\label{subsec:model_A1} 
For the system of two membranes linked by tethers (model E, cf.\ Fig.~\ref{fig:1}) 
the Hamiltonian consists of a bending energy term for each membrane 
and the potential energy of the $N$ tethers due to membrane elongations at the tether sites $\vec{r_\alpha}$, 
\begin{eqnarray}   \label{eqn:Def_H(2Mem)}
   H^{\scriptscriptstyle \mathrm{(f)}}  & \equiv &  
       \int_{0}^{L_x}  \!\!\!\!\! \d x   \int_{0}^{L_y}  \!\!\!\!\! \d y 
                 \left\{      \frac{\kappa_1}{2}  \big[\nabla^2 h^{\scriptscriptstyle (1)}(\vec{r}) \big]^2    
     +    \frac{\kappa_2}{2} 
                 \big[\nabla^2 h^{\scriptscriptstyle (2)}(\vec{r}) \big]^2  \right\}   \nonumber \\
 &&  +  {}\;  \sum\nolimits_{\alpha = 1}^{N} \,\frac{K_{\alpha}}{2}  \left[ h^{\scriptscriptstyle (1)}(\vec{r_{\alpha}}) 
                 - h^{\scriptscriptstyle (2)}(\vec{r_{\alpha}})  \right]^2    \ \ .
\end{eqnarray}
The involved quantities are explained in section~\ref{sec:2}. 
In order to avoid a divergence of the translational modes, we preliminarily exclude the translational mode of the lower membrane or, 
in other words, switch over to the frame of reverence given by the spatial mean height of the lower membrane 
(cf.\ subsection \ref{subsec:2M_conti}).

Inserting Fourier expansions for both membrane profiles, 
\begin{equation}  \label{eqn:Fourierreihe} 
   h^{\scriptscriptstyle (i)}(\vec{r})  ~ \equiv ~   
  \sum_\vec{k} \,h^{\scriptscriptstyle (i)}_{\vec{k}}\, e^{i\, \vec{k} \cdot \vec{r}}  
\end{equation}
with $i \in \{1, 2\}$, $\vec{k}  \equiv  (k_x, k_y)$ and $k_{x,y}  = (0, \pm 1, \pm 2, \ldots) \cdot (2 \pi / L_{x,y})$, 
and translating to independent variables \cite{Lin_Brown_BiophysJ_2004}, 
the Hamil\-ton\-ian can be expressed in the following form \cite{Merath_DA}, 
\begin{equation} \label{eqn:H=dMd}
   H ~ = ~ \vec{d}^\top  \vec{M} \, \vec{d}   
   ~ \equiv ~ \sum\limits_{t, t^\prime} d^{~}_t \:  M^{~}_{t, t^\prime}  \: d^{~}_{t^\prime}  \ .
\end{equation}
The vector $\vec{d}$ of independent system variables is built up by 
the vectors $\vec{c}^{\scriptscriptstyle (i)}$ of independent variables of both membranes, 
\begin{equation}
   \vec{d} ~ \equiv ~    
        \bigg(  \! \begin{array}{c} 
                   \vec{c}^{\scriptscriptstyle (1)}    \\[0.4ex] 
                   \vec{c}^{\scriptscriptstyle (2)} 
                   \end{array} \! \bigg)   \ .
\end{equation}
Their components $c^{\scriptscriptstyle (i)}_r$ (with $r = 0, 1, 2, \ldots$) can be grouped into three sectors, 
\begin{equation}
   {\vec{c}^{\scriptscriptstyle (i)} }^\top
   \equiv \Big( {\textstyle \frac{1}{2} }h^{\scriptscriptstyle (i)}_{\vec{0}}, \;   
            \{ \mathrm{Re}\,h^{\scriptscriptstyle (i)}_{\vec{q}} \}  , \; 
            \{ \mathrm{Im}\,h^{\scriptscriptstyle (i)}_{\vec{q}} \}  \Big)  \ , 
\end{equation}
where $\vec{q}$ runs through all independent, non-vanishing wave vectors
and $\mathrm{Re}$ and $\mathrm{Im}$ denote the real and imaginary part, respectively. 
Note that, if the translation mode of a membrane is manually excluded (to avoid its divergence), 
the corresponding vector and matrix elements have to be discarded.

The symmetric, positive definite matrix $\vec{M}$ consists of four quadrants and can be written as  
\begin{equation}
   \vec{M}  ~  \equiv  ~   
   \left( \begin{array}{cc} 
                \vec{M}^{\scriptscriptstyle (1)}  \hphantom{}   &   \!\! \vec{H}   \\[0.7ex]
                \!\! \vec{H}                                             &  \, \vec{M}^{\scriptscriptstyle (2)}
          \end{array} \!\!  \right)  \  ,
\end{equation}
if the set of possible wave vectors are the same for both membranes. 
The submatrices $\vec{M}^{\scriptscriptstyle (1)}$ and $\vec{M}^{\scriptscriptstyle (2)}$ represent a single membrane tethered to a
substrate~\cite{Merath_Seifert_PRE_2006}, 
whereas the matrix $\vec{H}$ contains the interaction of Fourier modes of both membranes. 
These matrices can be simplified to 
\begin{equation}   \label{eqn:Def_submatrix_Mi}
    M^{\scriptscriptstyle (i)}_{r,r^{\prime}} \, = \, E^{\scriptscriptstyle (i)}_r \cdot \delta^{~}_{r,r^{\prime}} 
                        \:+\: \sum_{\alpha = 1}^{N} \, m_r^{\alpha} \cdot m_{r^{\prime}}^{\alpha}  
\end{equation}
and 
\begin{equation}
    H^{~}_{r,r^{\prime}}  
    ~ = ~  - \,  \sum_{\alpha = 1}^{N} \, m_r^{\alpha} \cdot m_{r^{\prime}}^{\alpha} 
\end{equation}
with 
\begin{equation}
   E^{\scriptscriptstyle (i)}_r ~ \equiv ~ \left\{ \begin{array}{ll}    0              &  \quad \mathrm{for~}r=0  \\[0.6ex] 
                        \kappa_i \, L_x L_y \, |\vec{q}(r)|^4   &   \quad \mathrm{for~}r>0   \end{array} \right.   
\end{equation}
and  
\begin{equation}  \label{eqn:Def_m}
   m_r^{\alpha} \: \equiv \: \left\{ \begin{array}{ll} 
        \!\!\!        ~~\:   \sqrt{2 K_{\alpha}}      
                                               &  ~ \mathrm{for~} r=0 ~\mathrm{(1^{st}~sector)}  \hspace*{-0.5cm}   \\[0.7ex]
        \!\!\!    ~~\:    \sqrt{2 K_{\alpha}} \, \cos[\vec{q}(r) \cdot \vec{r}_{\alpha}]  &  
                                       ~ \mathrm{for~the~2^{nd}~sector}      \hspace*{-4cm}          \\[0.7ex]
                                       \!\!\! -\, \sqrt{2 K_{\alpha}} \, \sin[\vec{q}(r) \cdot \vec{r}_{\alpha}]  &  
                 ~ \mathrm{for~the~3^{rd}~sector} \ \, .
         \hspace*{-4cm}   \end{array}     \right.  
\end{equation}

The inverse matrix $\vec{M}^{-1}$ provides the correlations of all Fourier coefficients, 
\begin{equation}  \label{eqn:correlations_d_t} 
   \langle\, d_t \, d_{t^{\prime}} \,\rangle  
   ~ = ~   \frac{k_B T}{2}   \; \vec{M}^{-1}_{t,t^\prime}   \  ,
\end{equation}
thus in particular the fluctuation spectrum $\langle \, |h^{\scriptscriptstyle (i)}_\vec{k}|^2  \, \rangle$ 
of each membrane can be extracted.

\subsection{Model F: Additional solid support}
\label{subsec:solid_support} 
The extension to the membrane system additionally tethered to a solid support (model F, cf.\ Fig.~\ref{fig:1}) 
is straightforward.  
Let $N^\mathrm{(s)}$ vertical, harmonic tethers of strength $K^\mathrm{(s)}_{\alpha}$ 
at lateral sites $\vec{r^\mathrm{(s)}_{\alpha}}$ attach the lower membrane to the substrate. 
The corresponding Hamiltonian reads 
\begin{equation}    
   H^{\scriptscriptstyle \mathrm{(G)}}
   ~ \equiv ~
      H^{\scriptscriptstyle \mathrm{(F)}} 
        +   \sum\nolimits_{\alpha = 1}^{N^\mathrm{(s)}} \,\frac{K^\mathrm{(s)}_{\alpha}}{2}  
            \left[  h^{\scriptscriptstyle (2)}\big(\vec{r^\mathrm{(s)}_{\alpha}}\big)  \right]^2   \ .
\end{equation}

The additional term finds its corresponcance in the calculation scheme 
by a modification of the submatrix $\vec{M}^{\scriptscriptstyle (2)}$ (cf.\ equation (\ref{eqn:Def_submatrix_Mi})), 
\begin{equation}
    M^{\scriptscriptstyle (2)}_{r,r^{\prime}} \, = \, E^{\scriptscriptstyle (2)}_r \cdot \delta^{~}_{r,r^{\prime}} 
                        \:+\: \sum_{\alpha = 1}^{N} \, m_r^{\alpha} \cdot m_{r^{\prime}}^{\alpha}  
                        \:+\: \sum_{\alpha = 1}^{N^\mathrm{(s)}} \, n_r^{\alpha} \cdot n_{r^{\prime}}^{\alpha}  
\end{equation}
with $n_r^{\alpha}$ defined analogously to $m_r^{\alpha}$ (cf.\ equation (\ref{eqn:Def_m})) 
employ\-ing $K^\mathrm{(s)}_{\alpha}$ instead of $K_{\alpha}$, 
i.\,e., $n_r^{\alpha} \equiv m_r^{\alpha}|_{K_{\alpha} \to K^\mathrm{(s)}_{\alpha}}$.

If the solid support is built up by a continuous interacting potential 
instead (cf.\ the last term in equation (\ref{eqn:Def_H2co_s})), 
the associated parameter $\gamma$ must be added to $E^{\scriptscriptstyle (2)}_r$, i.\,e.,
\mbox{$E^{\scriptscriptstyle (2)}_r  \to  E^{\scriptscriptstyle (2)}_r + \gamma\,$}.

\subsection{Fluctuation spectra}
\label{subsec:spectra_discrete} 
Quasi-analytical results for two membranes (discretely) tethered between each other and possibly also to a substrate 
are presented and compared to the corresponding models with continuous interactions. 
First fluctuation spectra for the different cases are analyzed, then real-space elongations are examined.

In order to make the continuous harmonic potential comparable to an array of equally strong discrete tethers, 
we replace the stiffness parameter $\gamma$ by the parameters 
of the discrete model, that lead to the same effective spring constant for planar membrane elongations,   
\begin{equation}  \label{eqn:gamma_K_Delta}
   \gamma = \frac{1}{L_x L_y} \sum\nolimits_{\alpha = 1}^{N} K_\alpha = \frac{N \, K}{L_x L_y} = \frac{K}{\Delta^2} \  .
\end{equation}
Then the reduced wave vector $z$, used in the continuous cases, is related to $k \Delta$ via 
\begin{equation}
   z ~ = ~ \frac{k \Delta}{(K \Delta^2 / \kappa)^{1/4}}  \ .
\end{equation}

Except for the free membrane case (and pathological cases in the tether models), 
the fluctuation spectrum exhibits a relative extremum at the spectral origin. 
We do not count this trivial extremum and, in the following, consider extrema only for non-vanishing wave vectors.

\subsubsection{Fluctuation spectra of model E (without substrate)}
\label{subsubsec:spectra_modelA1} 
\begin{figure}[!t]
\hfil \hspace{-0.0cm}  \includegraphics[scale=0.59,clip]{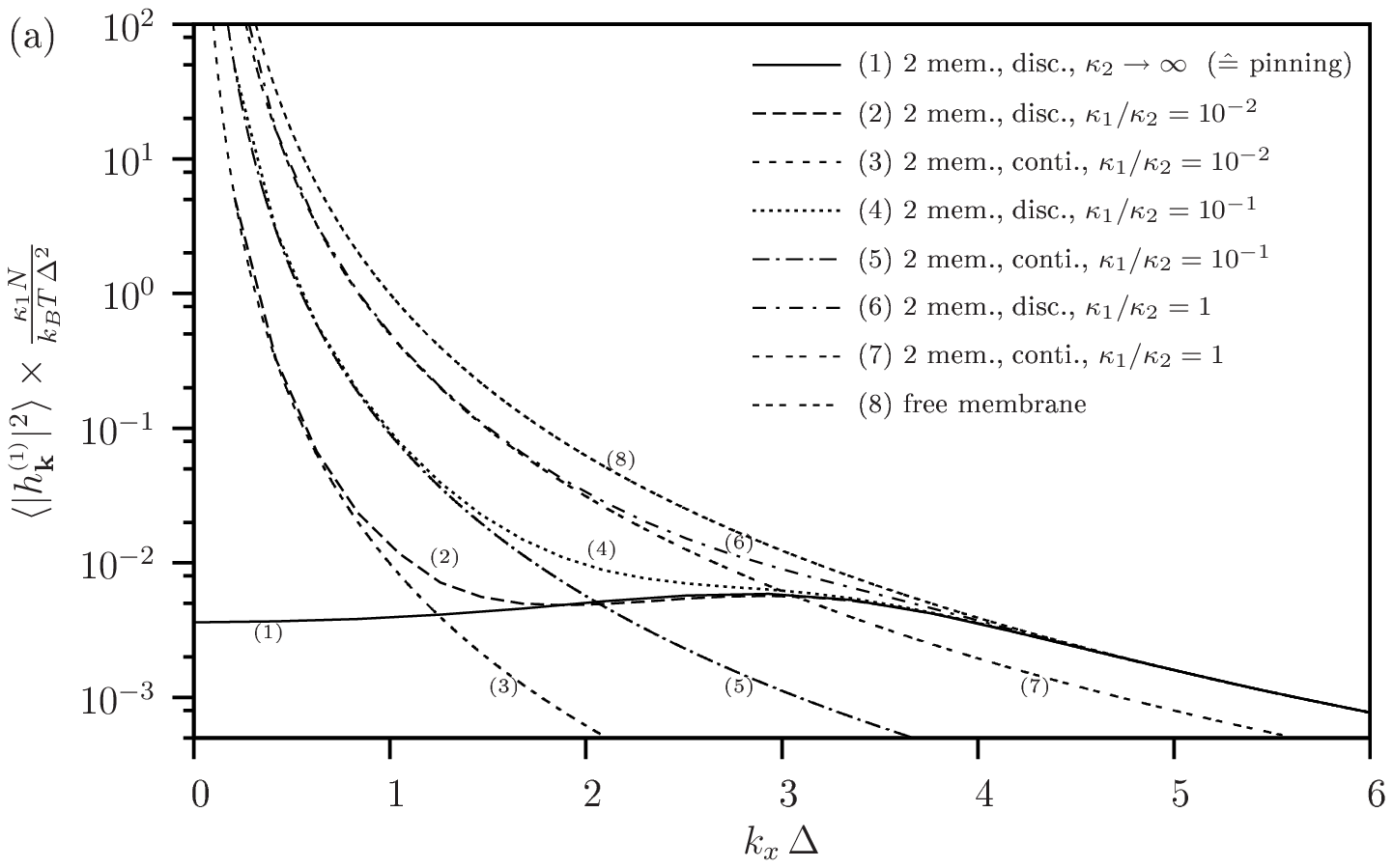}   \hfil

\vspace{0.2cm}
\hfil \hspace{-0.0cm}  \includegraphics[scale=0.59,clip]{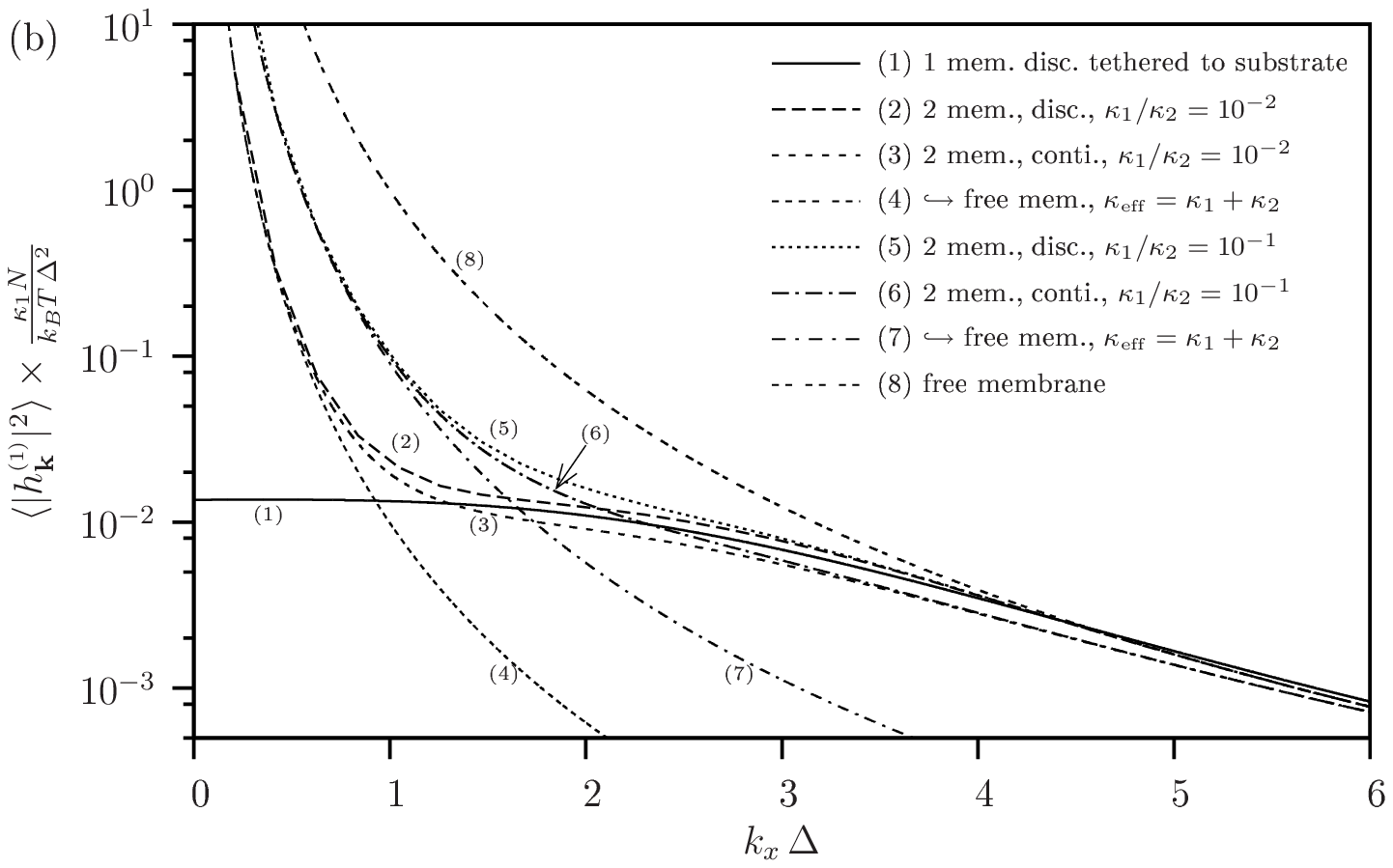}   \hfil
\caption{Fluctuation spectra for model E, i.\,e., of a membrane attached to a second (lower) quasi-parallel membrane 
         by a quadratic meshwork of equally stiff tethers.  
         The frame of reference is the spatial average of the lower membrane heigth profile, 
         thus all tethered membrane spectra in (a) and (b), respectively, coincide at the origin. 
         Cuts through the spectra along a main axis are displayed.
         \newline
         (a) Rigid tethers ($K \to \infty$). 
             The spectrum $\langle \, |h^{\scriptscriptstyle (1)}_{\vec{k}}|^2 \, \rangle$ 
             exhibits different regimes for varying values of $\kappa_1 / \kappa_2$. 
             A single membrane ``pinning'' spectrum is found for $\kappa_1 / \kappa_2 \to 0$.  
             For increasing ratio $\kappa_1 / \kappa_2$ 
             the relative minimum moves away from the origin and is less pronounced. 
             The spectrum remains nonmonotonic up to 
             $\kappa_1 / \kappa_2 \simeq 10^{-2}$ (curve 2), where a transition between nonmonotonicity and a monotonicity occurs. 
             A cross-over region between the (long wavelength) behavior as if the discrete tethers were a continuous potential 
             and the (short wavelength) behavior as if the lower membrane was completely stiff (like a substrate) 
             is situated around the intersection of both limiting curves. 
             The expression for the continuous interaction spectrum takes a simple form in the present case, 
             namely that of a free membrane with a modified bending rigidity. 
             Finally, for $\kappa_1 / \kappa_2 \to \infty$ the upper membrane fluctuates like a free membrane. 
         ---
         (b) Relatively strong tethers. 
             In the displayed case of $K \Delta^2 / \kappa_1 = 10^2$ the (two-dimensional) single membrane spectrum 
             is at the transition between monotonic and nonmonotonic behavior. 
             In comparison to the situation in (a), here the con\-ti\-nu\-ous interaction spectrum represents a better approximation 
             for wave vectors close to the cross-over region. 
} 
\label{fig:2freeM_discrete}
\end{figure}

Fluctuation spectra of a tethered membrane pair (model E, cf.\ Fig.~\ref{fig:1}) are considered. 
This compound membrane is mobile as a whole since there is no contact to a substrate. 
As a choice for the tethers connecting the two almost parallel membrane sheets, 
we consider a quadratic array of equally strong tethers, $K_\alpha \equiv K$. 
The translational mode of the lower membrane is excluded (cf.\ section~\ref{subsec:model_A1}), 
and we focus on the upper membrane and its spectrum $\langle \, |h^{\scriptscriptstyle (1)}_\vec{k}|^2  \, \rangle$. 
Several spectra belonging to model E are shown and compared to limiting cases in Fig.~\ref{fig:2freeM_discrete}. 

Tethered membrane spectra are anisotropic for large enough spring constants 
(cf.\ Fig.\ 2 in Ref.\ \cite{Merath_Seifert_PRE_2006}), 
while spectra associated to continuous interactions are isotropic. 
Since the spectrum for a membrane system mobile as a whole co\-vers several orders of magnitude in the interesting region, 
azimuthal averages are vulnerable to numerical artefacts. 
Thus the spectra in Fig.\ \ref{fig:2freeM_discrete} are evaluated along a main axis in Fourier space (e.\,g., $\vec{k}=(k_x,0)$),  
which is related to a main axis ($\vec{r}=(x,0)$) of the quadratic tether meshwork in real-space.

We distinguish between four different, partially overlapping regions in the parameter space 
constructed by the two parameters $\kappa_1/\kappa_2$ and $K \Delta^2/\kappa_1$: 
(i) a soft lower membrane, 
(ii) a rigid lower membrane,
(iii) weak tethers, 
and (iv) relatively stiff tethers and a stiff lower membrane. 

The relationship between the different membrane models (cf.\ Fig.\ \ref{fig:1}) reveals the following two trivial limiting cases. 
The case (i) of a soft lower membrane (i.\,e., $\kappa_1/\kappa_2 \to \infty$) corresponds to a free membrane 
(except at the origin, due to choice of the frame of reference). 
Conversely, for (ii) a rigid lower membrane (i.\,e., $\kappa_1/\kappa_2 \to 0$) 
the behavior of a single membrane attached to a substrate 
is recovered.

(iii) For weak tethers (i.\,e., for $K \Delta^2/\kappa_1 \ll 10^2$) 
the spectrum $\langle \, |h^{\scriptscriptstyle (1)}_\vec{k}|^2  \, \rangle$
is well approximated by the formula for two membranes jointed by a continuous potential 
(cf.\ section~\ref{subsec:2M_conti}). 
It is reasonable that the spectrum of a membrane tethered to a second membrane lies above 
the corresponding spectrum for attachment to a substrate (that complies with $\kappa_2 \to \infty$). 
Thus it is plausible that the continuous description cannot be valid for wave vectors for which the ``continuous spectrum'' 
lies below the single membrane ``pinning'' spectrum. 
The intersection of the continuous and the pinning spectrum marks a \emph{cross-over} region between a 
small wave vector region, where the \emph{continuous} formula is valid, 
and a large wave vector region, where the spectrum is well described by the \emph{pinning} spectrum. 
Thus for long-wave undulations the discreteness of the tether meshwork becomes less important, 
and for short-wave undulations the lower membrane acts like a substrate. 
Note that for $K \to \infty$ the continuous spectrum is equivalent for all wave vectors (cf.\ equation (\ref{eqn:limits_spectrum})) 
to the behavior of a single membrane with an effective bending rigidity 
$\kappa_{\scriptscriptstyle \mathrm{eff}} \equiv \kappa_1 + \kappa_2$.

The case (iv) of relatively stiff tethers and a stiff lower membrane comprises 
a \emph{nonmonotonic} (two-dimensional) spectrum $\langle \, |h^{\scriptscriptstyle (1)}_\vec{k}|^2  \, \rangle$ 
with a relative minimum and a relative maximum.  
The monotonic and the nonmonotonic regime are separated by a line in ($\kappa_1 / \kappa_2, K \Delta^2 / \kappa_1$)-space 
connecting two points,  
firstly ($\kappa_1 / \kappa_2 = c_1 \simeq  10^{-2}$, $K \Delta^2 / \kappa_1$ $\to \infty$) and    
secondly ($\kappa_1 / \kappa_2$ $\to \infty$, $K \Delta^2 / \kappa_1 = c_2 \simeq 10^2$). 
The nonmonotonic region is located on the side of this ``coexistence line'' which possesses smaller $\kappa_1 / \kappa_2$ 
or/and larger $K \Delta^2 / \kappa_1$ values. 
Hence, in other words, nonmonotonicity in model E 
occurs only for relatively stiff tethers and two membranes possessing very different bending rigidities.

All spectra for model E share the qualitative beha\-vior at the spectral origin ($\vec{k} = \vec{0}$)  
where the translational mode of the lower membrane has been excluded to avoid a divergence at the origin. 
Within numerical accuracy we can state that $\langle \, |h^{\scriptscriptstyle (1)}_\vec{0}|^2  \, \rangle$ 
for discrete tethers coincides with the spectrum of a single membrane tethered to a substrate. 
Thus, all discrete spectra in parts (a) and (b), respectively, of Fig.\ \ref{fig:2freeM_discrete} 
take the same value at the origin and then jump to a high value 
(which is related to the smallest wave vector larger than zero and therefore to the extensions of the basal plane). 
This can be understood by a simple consideration: 
For an equidistant meshwork of equally stiff tethers 
let any Fourier mode (except the a priori forbidden translational mode) of the lower membrane 
push the upper membrane away at some lateral position $(x,y)=(\tilde{x},\tilde{y})$. 
Then there is always a symmetric position $(x,y)=(L_x-\tilde{x}, L_y-\tilde{y})$, 
where a tether pulls in the other direction. 
Thus the net effect on the lateral spatial average 
(that equals $h^{\scriptscriptstyle (1)}_{\vec{0}}$) 
of the upper membrane height profile vanishes, i.\,e., 
undulations of the lower membrane do not influence 
$\langle \, |h^{\scriptscriptstyle (1)}_{\vec{0}}|^2 \, \rangle$.

\subsubsection{Fluctuation spectra of model F (with substrate)}
\label{subsubsec:spectra_modelA2} 
After the freely moving two-membrane system, we now investigate 
the model with additional solid support tethered 
to the lower membrane (model F, cf.\ Fig.~\ref{fig:1}). 
For simplicity we assume the additional sub\-strate--mem\-brane tether meshwork 
to be identical to the quadratic mem\-brane--mem\-brane tether meshwork, i.\,e., 
$N^\mathrm{(s)} = N$, $K^\mathrm{(s)}_{\alpha} = K$, 
and $\vec{r^\mathrm{(s)}_{\alpha}} = \vec{r^{~}_{\alpha}}$.

In the tethered membrane model the tether distance $\Delta$ imposes a natural length scale. 
In contrast to the continuous interaction model (cf.\ equation (\ref{eqn:Def_reduced_wavevector})), 
the spectra cannot be expressed in terms of a reduced wave vector that 
incorporates the spring constant.  
The spectrum of the upper or lower membrane can be expressed in terms of a scaling function $f^\mathrm{\scriptscriptstyle (i)}$ 
with $i \in \{1, 2\}$, 
\begin{equation}
   \langle \, |h^{\scriptscriptstyle (i)}_\vec{k}|^2 \, \rangle 
   ~ = ~ \frac{k_B T}{\kappa_i N} 
         \times  f^\mathrm{\scriptscriptstyle (i)} \bigg( \frac{\kappa_1}{\kappa_2}, 
                                                          \frac{K \Delta^2}{\kappa_i}, k_x\Delta, k_y\Delta  \bigg)   \  .
\end{equation}

\begin{figure}[!t]
\hfil \hspace{-0.0cm}  \includegraphics[scale=0.59,clip]{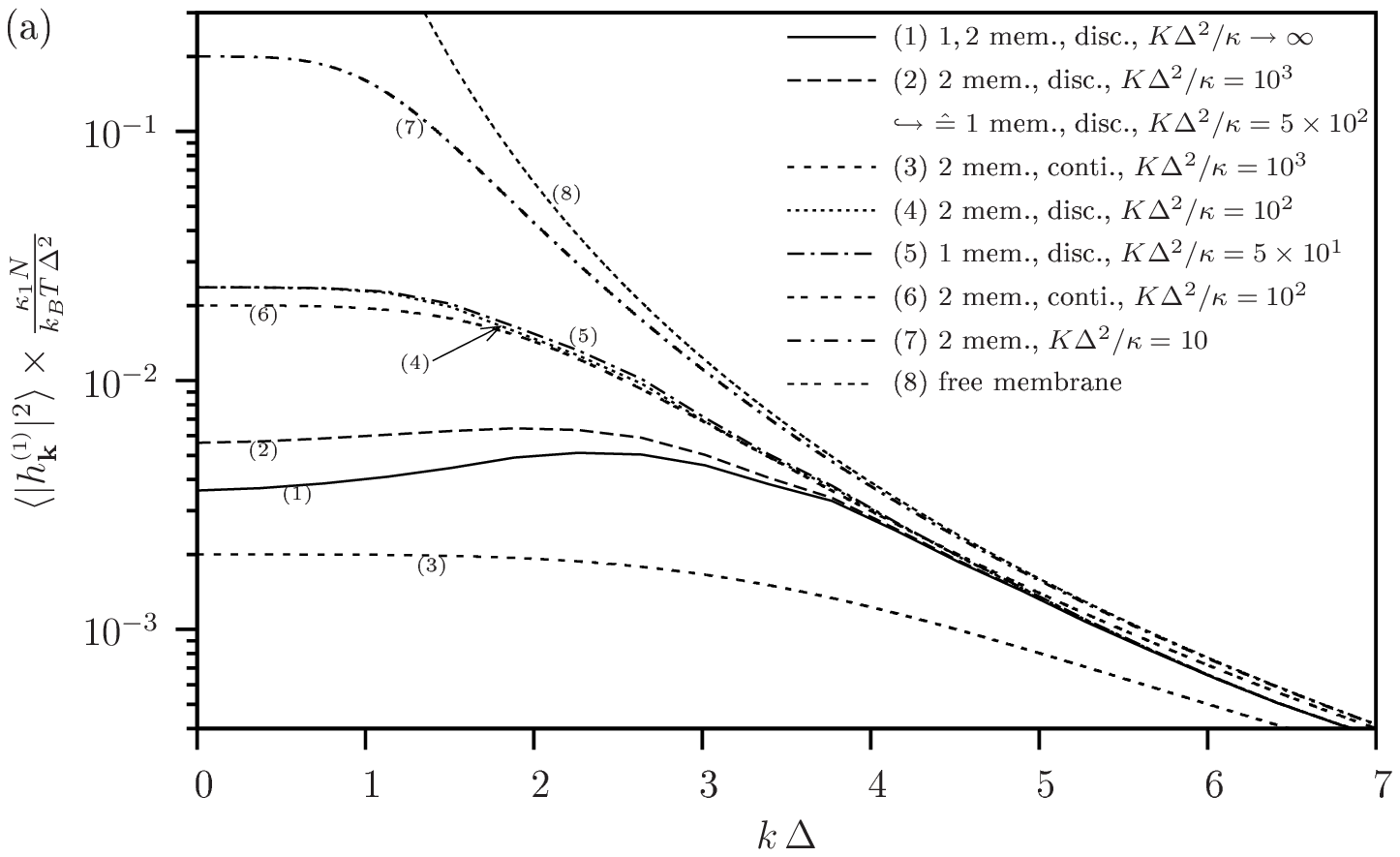}   \hfil

\vspace{0.2cm}
\hfil \hspace{-0.0cm}  \includegraphics[scale=0.59,clip]{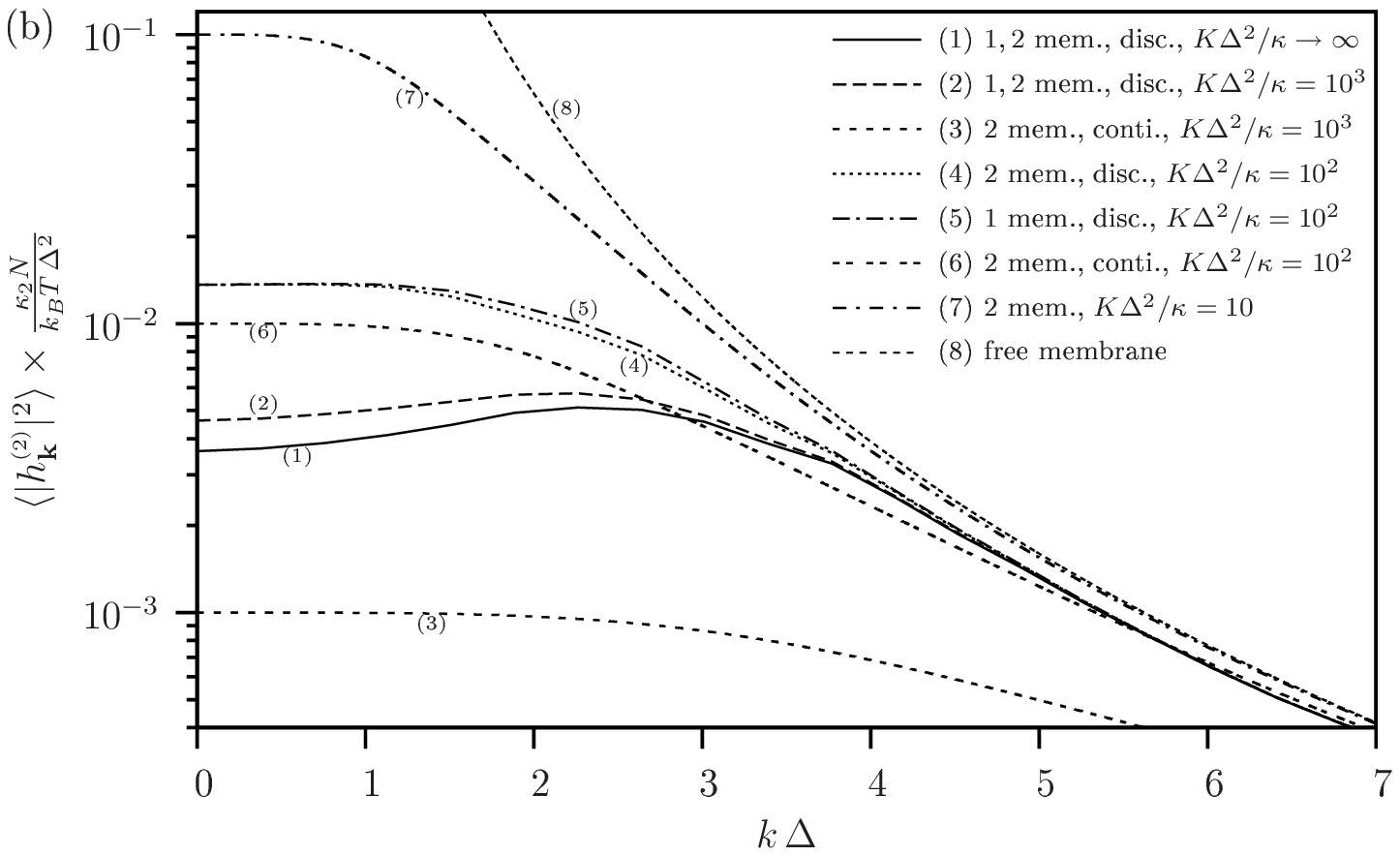}   \hfil
\caption{Fluctuation spectra of the upper (a) and lower (b) membrane in model F, 
         i.\,e., for two identical membranes ($\kappa_1=\kappa_2$) 
         tethered among each other and to a substrate. 
         Both tether meshworks form quadratic arrays of equally stiff tethers situated directly on top of each other. 
         The two-dimensional spectra are averaged with respect to the azimuthal angle in the $(k_x, k_y)$-plane 
         and plotted versus the radial part $k$.  \newline  
         (a) For stiff tethers the spectrum of the upper membrane behaves like a single membrane spectrum 
             with an effective spring constant $K_\mathrm{\scriptscriptstyle eff} = K/2$ (curves 1 and 2). 
             It converges to the spectrum of a two membrane model with continuous interaction for small spring constants 
             (curves 7 and 8). 
             In between, i.\,e., in the regime of intermediate values of the spring constant (curve 4), 
             slight deviations from those limiting cases (curves 5 and 6) appear for undulations modes with wavelengths 
             around a few tether distances. 
         ---
         (b) The lower membrane features the same spring constant regimes as the upper membrane, 
             but the limiting case of a single membrane (i.\,e., for stiff tethers)
             possesses the same spring constant ($K_\mathrm{\scriptscriptstyle eff} = K$) as the original tethers, 
             since for the lower membrane the two tether meshworks are arranged in parallel and not in series. 
} 
\label{fig:2M_discrete_topbottom}
\end{figure}

We proceed as follows. For both the upper and the lower membrane we distinguish three cases 
depending on the parameters $\kappa_1/\kappa_2$ and $K \Delta^2/\kappa_i$: 
(a) rigid tethers (i.\,e., $K~\to~\infty$),
(b) a vanishing bending rigidity of the other (i.\,e., not focussed) membrane, 
and (c) similar bending rigidities. 
Case (c) itself contains three regimes, 
(i) stiff tethers, 
(ii) intermediate spring constants, 
and (iii) weak tethers. 
Only the upper membrane exhibits an additional regime (d) for stiff tethers and a stiff lower membrane, 
where a rich nonmonotonic spectrum is found.

(a) For rigid tethers ($K \to \infty$)
the spectrum of a single membrane pinned to a substrate 
is recovered independently from the two bending rigidities 
due to the two tether meshworks situated directly on top of each other (in ``phase''). 

(b) 
For a small bending rigidity $\kappa_2$ of the lower membrane (i.\,e., large values of $\kappa_1/\kappa_2$) 
bending modes of the lower membrane are energetically favored. 
Therefore the upper membrane behaves like a single membrane attached at a substrate 
by discrete tethers with an effective spring constant $K_\mathrm{\scriptscriptstyle eff} = K/2$, 
since from the point of view of the upper membrane the two tether meshworks are aligned in series. 
Analogously for small $\kappa_1/\kappa_2$ 
the lower membrane behaves like a single membrane attached to a substrate 
by discrete tethers with spring constant $K$,  
since the lower membrane feels the two tether meshworks like arranged in parallel.

(c) For intermediate values of $\kappa_1/\kappa_2$ (i.\,e., $\kappa_1/\kappa_2 \simeq \mathcal{O}(1)\,$)  
the spectra of both membranes can be grouped into three regimes according to the tether stiffness. 
(i) For large spring constants ($K\Delta^2/\kappa_i \gtrsim 10^{3}$)  
each spectrum coincides with the discrete single membrane spectrum 
with an effective spring constant $K_\mathrm{\scriptscriptstyle eff}$. 
Focussing on the upper membrane necessitates $K_\mathrm{\scriptscriptstyle eff} = K/2$, 
while for the lower membrane $K_\mathrm{\scriptscriptstyle eff} = K$ applies. 
Consequently, a nonmonotonic spectrum of the upper or lower membrane, respectively, 
is observed for $K_\mathrm{\scriptscriptstyle eff} \Delta^2/\kappa_i \gtrsim 10^2$~\cite{Merath_Seifert_PRE_2006}. 
(ii) For an intermediate tether stiffness small, suppressive deviations from the corresponding single membrane spectrum 
appear around $k\Delta \simeq \pi$. 
Finally, for (iii) weak tethers ($K\Delta^2/\kappa_i \lesssim 10$) 
the formulae for a supported membrane pair with con\-ti\-nu\-ous interactions 
(cf.\ subsection~\ref{subsec:2M_substrate_conti}) hold. 
Spectra of the upper and lower membrane for $\kappa_1/\kappa_2 = 1$ 
for the above mentionned three spring constant regimes are displayed 
and compared to the limiting cases in Fig.~\ref{fig:2M_discrete_topbottom}.

\begin{figure}[!t]
\hfil \hspace{-0.1cm}  \includegraphics[scale=0.58,clip]{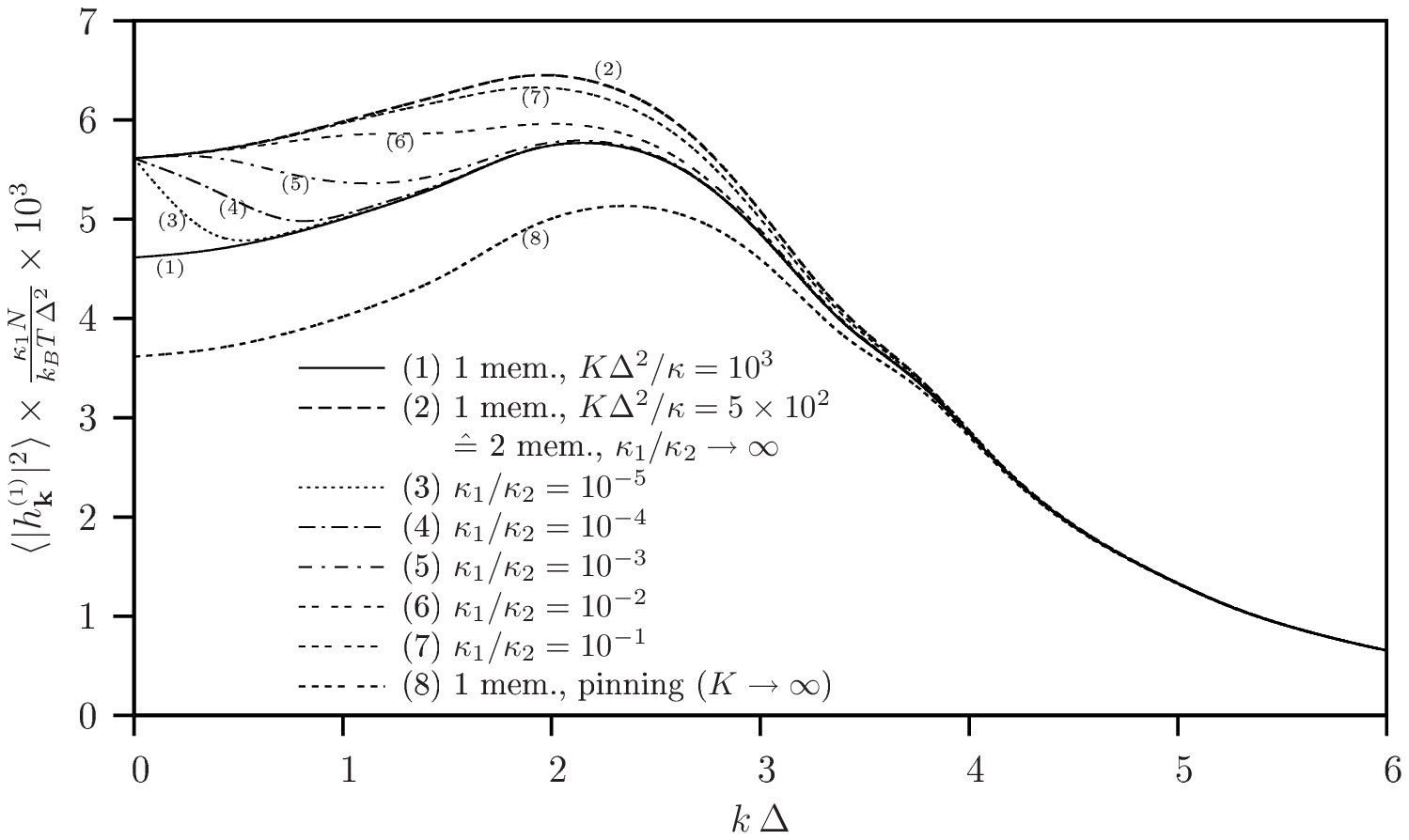}   \hfil
\caption{Nonmonotonic fluctuation spectra for unlike bending rigidities in model F.
         Two relative extrema (apart from the origin) appear in the spectrum of the upper membrane 
         in model F (cf.\ Fig.~\ref{fig:1}), if the lower membrane and the tethers are relatively stiff. 
         The new relative minimum for small wave vectors has evolved 
         from the relative minimum at the origin of a nonmonotonic single membrane spectrum, 
         since long wavelength undulations of the lower membrane enlarge the upper membrane's spectrum around the origin. 
         Azimuthal averages of the spectrum of the upper membrane are shown 
         for relatively stiff tethers ($K \Delta^2/\kappa_1 = 10^3$) 
         and various bending rigidity ratios $\kappa_1/\kappa_2$. 
         For comparison, also related spectra of single tethered membranes (curves 1,2, and 8) are displayed. 
         For $\kappa_1/\kappa_2 \to 0$ the spectrum resembles that of a single tethered membrane with the same spring constant
         (except at the origin). 
         For $\kappa_1/\kappa_2 \to \infty$ the spectrum of a single tethered membrane with half the spring constant
         is recovered. 
         For any value of $\kappa_1/\kappa_2$ the spectrum is situated between these two limiting cases, 
         but always maintains the same value at the origin. 
         The structure at $k \Delta \simeq \pi$ is an artefact from the azimuthal averaging 
         and does not appear in the original, two-dimensional spectra. 
} 
\label{fig:Nonmonotonicity_second_kind}
\end{figure}

(d) A qualitatively new effect arises for stiff tethers 
(at the order of the critical spring constant $\left. {K \Delta^2 / \kappa  }\right|_\mathrm{crit} \, \simeq \, 100 $, 
at which a single tethered membrane possesses a nonmonotonic spectrum)  
and for a relatively stiff lower membrane: 
Then the upper membrane behaves like a single tethered membrane (with spring constant $K$ and not $K/2$) 
only for intermediate and large wave vectors. 
Moreover, for small wave vectors the --- despite of the large $\kappa_2$ --- active translational mode of the lower membrane 
affects the upper membrane 
and thus increases the spectrum around the origin. 
For a sufficiently stiff lower membrane ($\kappa_1/\kappa_2 \lesssim 10^{-3}$) 
the spectrum of the upper membrane exhibits a \emph{relative minimum} at a small wave vector 
in addition to the familiar relative maximum. 
This new type of nonmonotonicity is shown in Fig.~\ref{fig:Nonmonotonicity_second_kind}. 
In the case of a single tethered membrane with stiff tethers a relative minimum is localized at the origin, 
but it moves to a non-vanishing wave vector for two membranes and suitable parameter values. 
In the limiting case of $\kappa_1/\kappa_2 \to \infty$ the spectrum of the upper membrane 
looks like a single tethered membrane spectrum except at the origin where the spectrum jumps to a higher value. 
Of course, very strong tethers hide this effect and yield the well-known pinning spectrum.

After the detailed discussion of the various regimes we qualitatively illustrate 
how the two membranes influence each other and lead to deviations from the behavior of a single tethered membrane. 
The influence of the lower on the upper membrane can be understood as follows: 
Undulations of the lower membrane displace the attachment sites of the upper tethers. 
With increasing bending elasticity of the lower membrane 
the probability of these attachment site fluctuations decreases. 
The concept of an effective spring constant only works well for negligible suppression of such displacements, 
so that two tethers on top of each other can be regarded as two springs in series. 
Since the effective (in general wave-vector-dependent) spring constant is smaller than $K$ 
(and equal to $K/2$ in the limiting case of $\kappa_2 \to 0$), 
the spectrum of the upper membrane always exceeds the corresponding single membrane spectrum. 

On the other hand, the influence of the upper on the lower membrane is slightly different: 
Here the limiting case for vanishing bending rigidity of the other membrane corresponds to a single tethered membrane spectrum with 
an effective spring constant equal to $K$. 
From the point of view of the lower membrane the two tether meshworks are arranged in parallel. 
A non-vanishing bending rigidity $\kappa_1$ leads to an effective (in general wave-vector-dependent) spring constant 
larger than $K$, thus the spectrum of the lower membrane alwalys stays below the corresponding single membrane spectrum.

\subsection{Real-space fluctuations}
\label{subsec:MSD_discrete} 
For a calculation of the mean-square displacement (MSD) of membrane $i$, 
correlations of the Fourier coefficients are necessary (cf.\ equation (\ref{eqn:MSD_general})). 
Since the sys\-tem with discrete tethers is not translationally invariant, 
all correlations have to be taken into account, not only those for $\vec{k}$ and $\vec{k}^\prime = -\vec{k}$ 
that comply with the fluctuation spectrum $\langle \, |h^{\scriptscriptstyle (i)}_\vec{k}|^2   \, \rangle$.

The height profile of membrane $i$ ($i = 1,2$) can be expressed as 
\begin{equation}
   h^{\scriptscriptstyle (i)}(\vec{r})  ~ = ~ 2 \, \vec{w}(\vec{r}) \cdot \vec{c}^{\scriptscriptstyle (i)}
\end{equation}
by means of 
\begin{equation}
   { \vec{w}(\vec{r}) }^\top ~ \equiv ~ ( 1, \{ \cos(\vec{q} \cdot \vec{r}) \},  \{ -\sin(\vec{q} \cdot \vec{r}) \}   )   \ .
\end{equation}
Then the MSD can be readily formulated as 
\begin{eqnarray} 
  \langle \, \big[h^{\scriptscriptstyle (i)}(\vec{r})\big]^2 \, \rangle & = &
             4 \, \sum\limits_{r, r^\prime} w^{~}_r(\vec{r}) \: 
                                            \langle \, c^{\scriptscriptstyle (i)}_{r^{\vphantom{\prime}}} \: 
                                                       c^{\scriptscriptstyle (i)}_{r^\prime}  \rangle \:
                                            w^{~}_{r^\prime}(\vec{r}) \nonumber  \\
    & = & 2 \; k_B T \; \sum\limits_{r, r^\prime} w^{~}_r(\vec{r}) \:  M^{-1}_{t(r,i), t(r^\prime,i)}  
                                               \: w^{~}_{r^\prime}(\vec{r}) \, , ~~~~
\end{eqnarray}
where equation (\ref{eqn:correlations_d_t}) was used. 
Herein $t(r,i)$ and $t(r^\prime,i)$ indicate the components --- belonging to the wave vectors labelled by $r$ and $r^\prime$ ---
of the appropriate quadrant of the matrix $\vec{M}^{-1}$, 
i.\,e., the upper left submatrix for $i = 1$ or the lower right submatrix for $i = 2$ (cf.\ subsection \ref{subsec:model_A1}).

\begin{figure}[!t]
\hfil \hspace{-0.1cm}  \includegraphics[scale=0.59,clip]{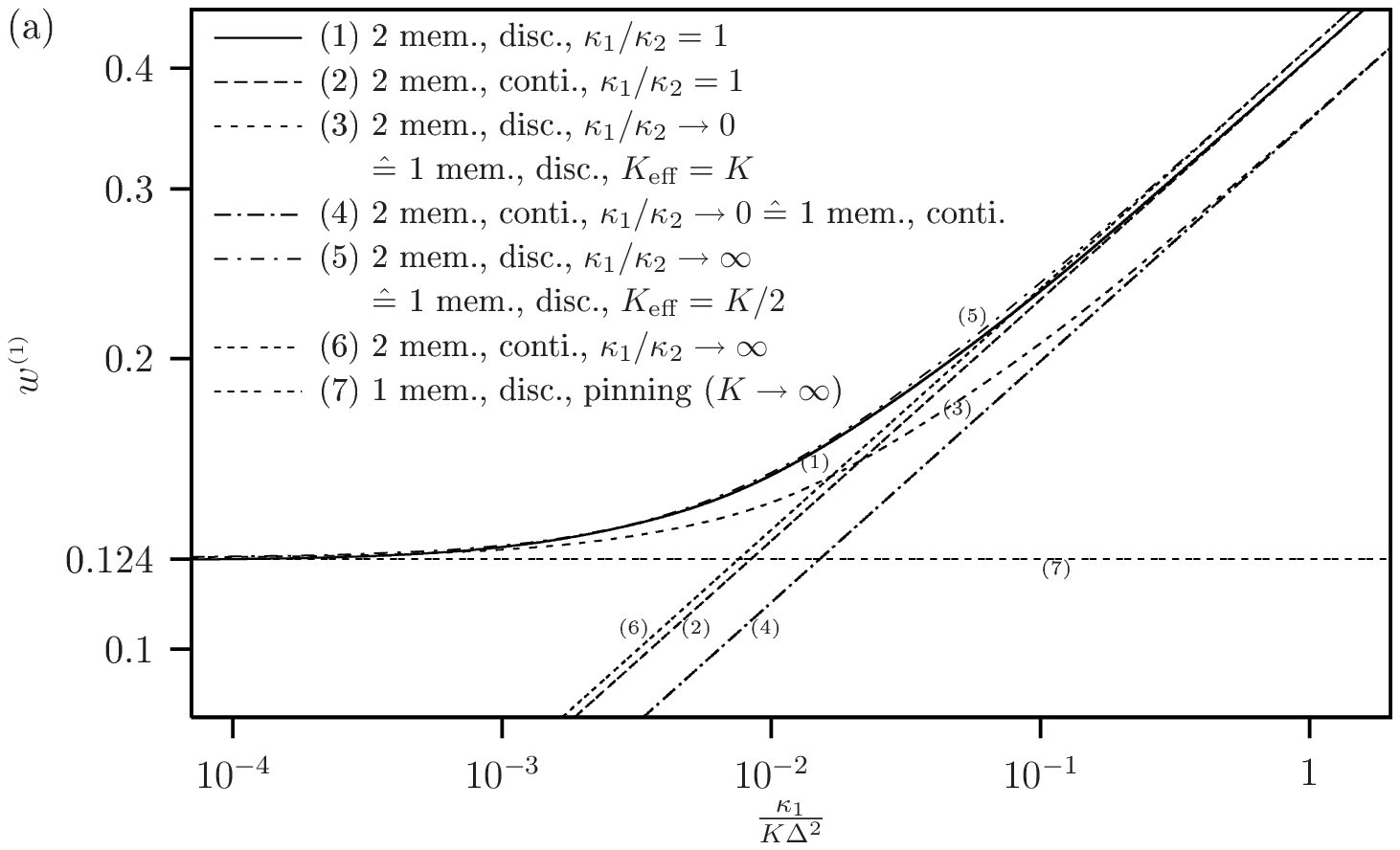}   \hfil

\hfil \hspace{-0.1cm}  \includegraphics[scale=0.59,clip]{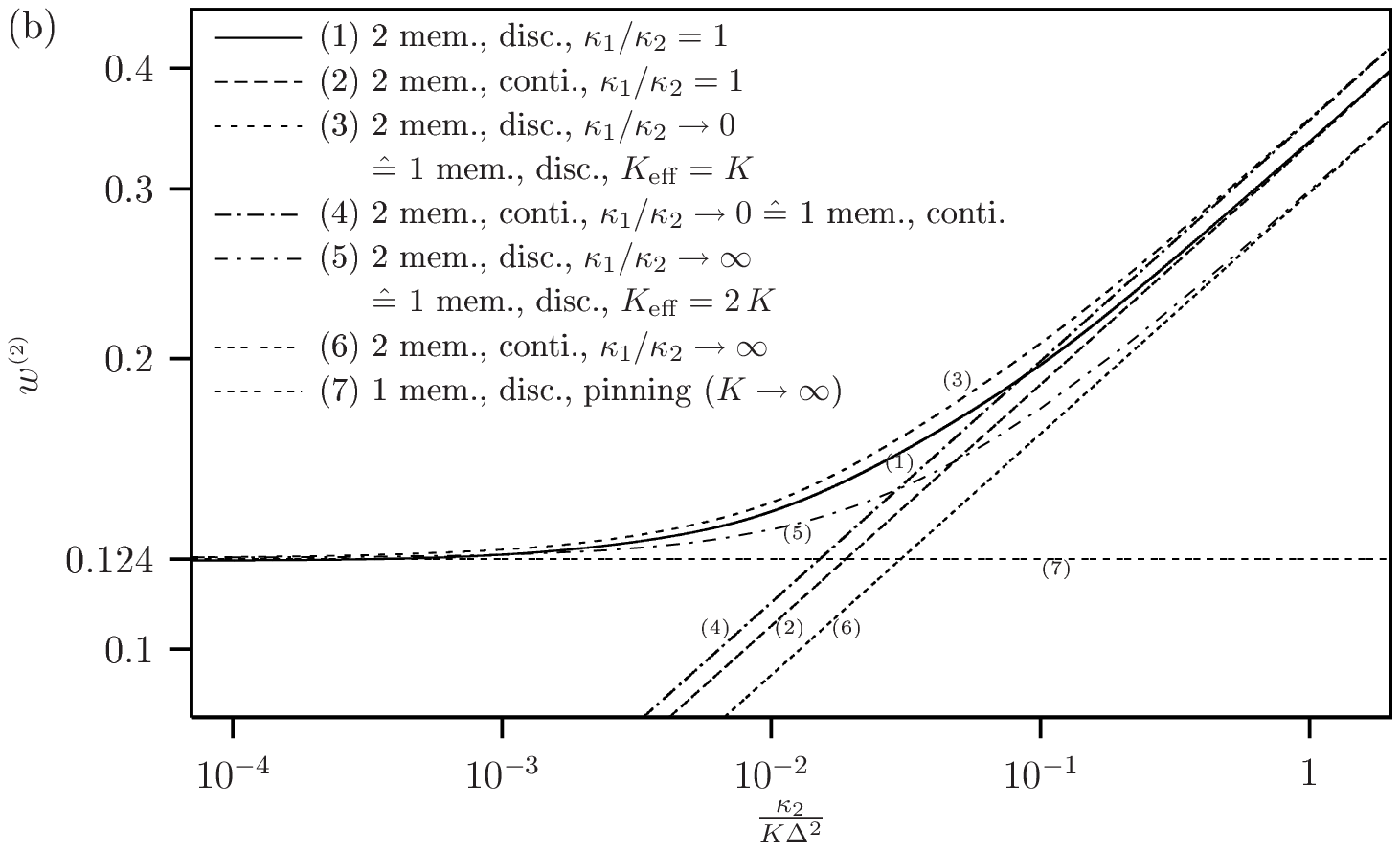}   \hfil
\caption{Typical displacements of the two membranes in model F (cf.\ Fig.~\ref{fig:1}). 
         The scaling functions $w^{\scriptscriptstyle (1)}$ (a) and $w^{\scriptscriptstyle (2)}$ (b) are displayed. 
         $w^{\scriptscriptstyle (i)}$ is related to the spatial maximum of the RMSD for membrane $i$ via equation~(\ref{eqn:Def.w}). 
         It is plotted versus the inverse of the dimensionless spring constant $K \Delta^2/\kappa_i$ 
         for some values of the ratio $\kappa_1/\kappa_2$ of the bending rigidities.  \newline
         (a): The curve for $w^{\scriptscriptstyle (1)}$ is situated between its two limiting cases. 
              For $\kappa_1/\kappa_2 \to 0$ the case of single tethered membrane with the same spring constant is recovered, 
              while for $\kappa_1/\kappa_2 \to \infty$ the upper membrane in model F behaves 
              like a single tethered membrane with half the spring constant. 
              For very stiff tethers $w^{\scriptscriptstyle (1)}$ converges to the value for a single pinned membrane, 
              $\simeq 0.124$. 
              For weak tethers the discretely tethered system approaches the behavior of the corresponding continuous model. 
         --- 
         (b): The above statements also hold analogously for $w^{\scriptscriptstyle (2)}$ of the lower membrane, 
              except that for $\kappa_1/\kappa_2 \to \infty$ the lower membrane behaves 
              like a single tethered membrane with twice the spring constant. 
              --- 
              Curves (3), (4), and (7) are identical in (a) and (b). 
              The RMSD of a single tethered membrane acts as a lower bound for the RMSD of the upper membrane in model F 
              and as an upper bound for the RMSD of the lower membrane. 
} 
\label{fig:RMSD_2M_disc_subs}
\end{figure}

Aiming at predictions for an experimental implementation of model F, we consider the root-mean-square displacement (RMSD) 
only for this model, i.\,e., for two membranes tethered to a substrate (cf.\ Fig.~\ref{fig:1}). 
The two tether meshworks are assumed to consist of two quadratic arrays of equally strong tethers 
situated directly on top of each other. 
The RMSD depends on the position in a unit cell of the tether meshwork (cf.\ Fig.\ 5 in Ref.~\cite{Merath_Seifert_PRE_2006}).

For the system at hand, the maximum of the RMSD appears in the middle of each unit cell and takes the form 
\begin{equation}   \label{eqn:Def.w}
   \mathrm{max}\left\{ \sqrt{\langle  \big[h^{\scriptscriptstyle (i)}(\vec{r})\big]^2 \rangle \, } \, \right\} 
             \equiv  { \sqrt{\frac{k_B T}{\kappa_i}\,} \Delta }
               \:   \times \:  w^{\scriptscriptstyle (i)}\bigg( \frac{K \Delta^2}{\kappa_i}, \frac{\kappa_1}{\kappa_2} \bigg)
\end{equation}
with a scaling function $w^{\scriptscriptstyle (i)}$ for the upper ($i=1$) or lower membrane ($i=2$). 
These two functions are presented in Fig.~\ref{fig:RMSD_2M_disc_subs}. 
They exhibit the following properties
in the limiting cases of rigid (i) or very weak (ii) tethers and in the intermediate regime (iii): 

(i) For rigid tethers, $K \to \infty$, the behavior of a single pinned membrane is recovered for both membranes. 
Note that this statement only holds if the two involved tether meshworks are identical and situated directly on top of each other.

(ii) For very weak tethers, $K \to 0$,  $w^{\scriptscriptstyle (i)}$ approaches the 
result for continuous interactions, i.\,e.,
\begin{equation}
   w^{\scriptscriptstyle (i)} ~ \to ~ 
   \frac{1}{\sqrt{8\,}}   \bigg( \frac{\kappa_i}{K \Delta^2}\bigg)^{\frac{1}{4}}   
                          \sqrt{b^{\scriptscriptstyle (i)}(\kappa_1/\kappa_2)}
\end{equation}
(cf.\ equations (\ref{eqn:MSD_2Mc_1}) and (\ref{eqn:MSD_2Mc_2}) and Fig.\ \ref{fig:2M_substrate_conti_MSD}). 
If additionally $\kappa_1/\kappa_2 \to 0$ 
(i.\,e., $\kappa_2 \to \infty$ if the upper membrane or $\kappa_1 \to 0$ if the lower membrane is considered),
the behavior of a single membrane continously attached to a substrate (with the same spring constant) 
is recovered for both membranes. 
If conversely $\kappa_1/\kappa_2 \to \infty$, 
both membranes act like a single, continuously interacting membrane, but 
with an effective spring constant $K_\mathrm{eff} = K/2$ for the upper membrane (since two tether meshworks are arranged in series)
and an effective spring constant $K_\mathrm{eff} = 2\,K$ for the lower membrane 
(since two tether meshworks --- one above and one below --- are arranged in parallel).

(iii) For intermediate values of the spring constant a cross-over between the two limiting cases 
(i.\,e., rigid and very weak tethers) 
occurs around the intersection of the limiting curves. 
In the example of equal bending rigidities, $\kappa_1=\kappa_2$, 
the cross-over is situated around $K \Delta^2/\kappa_1 \simeq 119$ for the upper membrane and 
around $K \Delta^2/\kappa_2 \simeq 53$ for the lower membrane. 
Thus, the cross-over roughly remains in the vicinity of the cross-over for a single tethered membrane, 
$K \Delta^2/\kappa \simeq 66$~\cite{Merath_Seifert_PRE_2006}.

\section{Conclusion and outlook}
\label{sec:5}
A formalism for determining the fluctuation spectrum and the root-mean-square displacement (RMSD) 
of fluid membranes coupled by discrete or con\-tinu\-ous interactions has been presented. 
Also hybrid models comprising discrete \emph{and} continuous interacting potentials are tractable: 
Firstly, if connections between two adjacent membrane sheets are built up by both discrete and 
continuous interacting potentials, 
the continuous interaction would lead to additional terms involving the harmonic po\-ten\-tial para\-meter $\gamma$. 
Secondly, membrane sheets with solely continuous interaction to neighboring objects 
could be dropped in the computation in favor of effective elastics constants 
for the neigh\-boring mem\-brane(s). 

The model of membrane pairs connected by \emph{discrete} tethers and optionally tethered to a substrate 
exhibit rich fluctuation spectra in comparison to the corresponding systems containing \emph{continuous} interacting potentials. 
In certain model configurations a nonmonotonic spectrum with \emph{two relative extrema} 
(cf.\ Figs.\ \ref{fig:2freeM_discrete} and \ref{fig:Nonmonotonicity_second_kind}) 
or with one relative extremum (cf.\ Fig.\ \ref{fig:2M_discrete_topbottom})
(in addition to the prevalent relative extremum at the spectral origin) was discovered.

The calculated RMSD serves as an estimate of the typical membrane elongations. This gives usefull hints to experimentalists, 
how the spacings (i.\,e., the rest-length of the tethers) between the membranes or/and the substrate 
have to be chosen in order to avoid undesired contact between them, 
e.\,g., if membrane-proteins are present which should not adhere to the substrate. 
Already in the \emph{continuous} interaction 
model the RMSD of a single tethered membrane can be increased or decreased 
by putting a second membrane below or above, respectively. 
Assuming equally strong interacting potentials, 
for identical bending rigidities ($\kappa_1 = \kappa_2$) the relative change of the RMSD is $-0.05$ or $+0.16$ 
and for an extremely soft lower membrane ($\kappa_1/\kappa_2 \to \infty$) $-0.16$ or $+0.19$ 
for the lower or upper membrane, respectively 
(cf.\ Fig.~\ref{fig:2M_substrate_conti_MSD}). 
Clearly, an upper interacting potential stronger than the lower one leads to more pronounced changes of the RMSD. 
While the stiffness parameter $\gamma$ of two equal continuous interacting potentials 
does not influence the relative change in the RMSD, 
the spring constant of tethers plays a decisive role here (cf.\ Fig.~\ref{fig:RMSD_2M_disc_subs}). 
In short, a two-membrane system could be utilized to modify typical membrane elongations to a certain extend.

The calculation presented here for two and three interacting membranes 
can be generalized to stacks of membranes~\cite{Merath_Seifert_in_preparation}. 
How different tether meshworks (e.\,g., a mesh\-work shif\-ted laterally with respect to the lower one)   
influence the fluctuations could be worth analyzing in future studies. 
Likewise, rather than looking at an effectively harmonic potential 
it would be interesting to include non-linear interactions describing 
steric interaction, van der Waals interaction, contributions due to hydration, 
and possibly electrostatic interaction~\cite{Lipowsky_Leibler_1986,Mecke_Charitat_Langmuir_2003}.
Finally, accounting for laterally mobile tether attachment sites in real biomembranes remains another challenge.

\vspace{0.4cm}
R.-J.\,M.\ thanks R.\ Richter  
and G.\ Majer  
for helpful discussions concerning experimental aspects.

\renewcommand{\theequation}{\Alph{section}.\arabic{equation}}
\setcounter{equation}{0}
\section*{Appendix}
\appendix

\section*{Three membranes connected by continuous harmonic potentials}
\label{subsec:3M_conti} 
The Hamiltonian for two continuously interacting membranes (model B, cf.\ equation (\ref{eqn:Def_H(M2,conti)})) 
can easily be extended to three membranes, 
\begin{eqnarray}   \label{eqn:Def_H(M3,conti)}
   H^{\scriptscriptstyle \mathrm{(3)}}  & \equiv &  
       \int_{0}^{L_x}  \!\!\!\!\! \d x   \int_{0}^{L_y}  \!\!\!\!\! \d y 
                \Big\{    
                \sum\nolimits_{i=1}^{3}  \frac{\kappa_i}{2}  \big[\nabla^2 h^{\scriptscriptstyle (i)}(\vec{r}) \big]^2  
                \nonumber \\
 &&   + {} \sum\nolimits_{i=1}^{2} 
             \frac{\gamma_i}{2}  \big[ h^{\scriptscriptstyle (i)}(\vec{r_{\alpha}}) 
                       - h^{\scriptscriptstyle (i+1)}(\vec{r_{\alpha}})  \big]^2  \Big\}  \  .  ~
\end{eqnarray}
A Fourier ansatz for the three heigth profiles (cf.\ equation (\ref{eqn:Fourier_ansatz})), 
separation in real and imaginary parts (cf.\ equation (\ref{eqn:real_and_imaginary})) 
and transformation to independent variables (cf.\ equation (\ref{eqn:independent_2co})) 
leads to 
\begin{eqnarray}  \label{eqn:independent_3co}
   H^{\scriptscriptstyle \mathrm{(3)}}  & = &  
      L_x L_y \sum\nolimits_{\vec{q}} \Big[
         ( \kappa_1 |\vec{q}|^4 + \gamma_1 ) \Big( {a_\vec{q}^{\scriptscriptstyle (1)}}^2  
             + {b_\vec{q}^{\scriptscriptstyle (1)}}^2 \Big)   \\ 
   &&   + ( \kappa_2 |\vec{q}|^4 + \gamma_1 + \gamma_2 ) 
                       \Big( {a_\vec{q}^{\scriptscriptstyle (2)}}^2  + {b_\vec{q}^{\scriptscriptstyle (2)}}^2 \Big)      \nonumber \\
   &&   + ( \kappa_3 |\vec{q}|^4 + \gamma_2 ) \Big( {a_\vec{q}^{\scriptscriptstyle (3)}}^2  
        + {b_\vec{q}^{\scriptscriptstyle (3)}}^2 \Big)
                            \nonumber \\
   &&   - 2 \gamma_1  \big( a_\vec{q}^{\scriptscriptstyle (1)} a_\vec{q}^{\scriptscriptstyle (2)} 
                        + b_\vec{q}^{\scriptscriptstyle (1)}  b_\vec{q}^{\scriptscriptstyle (2)} \big)  
        - 2 \gamma_2  \big( a_\vec{q}^{\scriptscriptstyle (2)} a_\vec{q}^{\scriptscriptstyle (3)} 
                        + b_\vec{q}^{\scriptscriptstyle (2)}  b_\vec{q}^{\scriptscriptstyle (3)} \big)  \Big]            \nonumber \\
   &&   + \frac{1}{2} L_x L_y \big[  \gamma_1 {a_\vec{0}^{\scriptscriptstyle (1)}}^2  
                                   + (\gamma_1 + \gamma_2) {a_\vec{0}^{\scriptscriptstyle (2)}}^2  
                                   - 2 \gamma_1 a_\vec{0}^{\scriptscriptstyle (1)}  a_\vec{0}^{\scriptscriptstyle (2)}     \big]  \ , \nonumber
\end{eqnarray}
where the translational mode $a_\vec{0}^{\scriptscriptstyle (3)}$ of the lowest membrane has been excluded to avoid divergences.

Pursuing the procedure of subsection \ref{subsec:2M_conti}, 
degrees of freedom of two membranes are integrated out, 
resulting in an effective Hamiltonian for the component membrane of interest. 
First, the degrees of freedom of the lowest (\mbox{$i=3$}) membrane are integrated out, 
\begin{eqnarray} 
   e^{-\beta H_\mathrm{eff}^{\scriptscriptstyle (1,2)}} 
      & \equiv &  \int {\mathcal D}h^{\scriptscriptstyle (3)} e^{- \beta H^{\scriptscriptstyle \mathrm{(3)}}}  \\
      & = & \bigg( \prod\nolimits_{\vec{q}} 
          \int_{- \infty}^{\infty}  \!\!\!\!\! \d a_\vec{q}^{\scriptscriptstyle (3)}  
          \int_{- \infty}^{\infty}  \!\!\!\!\! \d b_\vec{q}^{\scriptscriptstyle (3)}  \bigg) 
          e^{- \beta H^{\scriptscriptstyle \mathrm{(3)}}}   \ .  ~~
\end{eqnarray}
The effective Hamiltonian for the two upper membranes --- after a trivial shift of the energy scale --- then reads 
\begin{eqnarray}  \label{eqn:Def_H_1,2}
  && H_\mathrm{eff}^{\scriptscriptstyle (1,2)}
   =   L_x L_y \Big[  \big( \kappa_1 |\vec{q}|^4 + \gamma_1 \big) 
                        \big( {a_\vec{q}^{\scriptscriptstyle (1)}}^2  +  {b_\vec{q}^{\scriptscriptstyle (1)}}^2 \big)   \\
   &&                +  \big( \kappa_2 |\vec{q}|^4 + \gamma_1 + \gamma_2   -   \frac{\gamma_2^2}{\kappa_3 |\vec{q}|^4 + \gamma_2}   \big) 
                        \big( {a_\vec{q}^{\scriptscriptstyle (2)}}^2  +  {b_\vec{q}^{\scriptscriptstyle (2)}}^2 \big)   \nonumber  \\
   &&   - 2 \gamma_1  \big( a_\vec{q}^{\scriptscriptstyle (1)} a_\vec{q}^{\scriptscriptstyle (2)} 
                        + b_\vec{q}^{\scriptscriptstyle (1)}  b_\vec{q}^{\scriptscriptstyle (2)} \big)      \Big] 
      + \frac{1}{2} L_x L_y \big[ \ldots \mathrm{\scriptstyle (cf.\ equation\ (\ref{eqn:independent_3co}))} ~   \big]   . \nonumber
\end{eqnarray}

Now we can focus on the upper membrane ($i=1$) and integrate out the independent variables 
belonging to the middle membrane ($i=2$), 
\begin{eqnarray} 
   e^{-\beta H_\mathrm{eff}^{\scriptscriptstyle (1)}} 
      & \equiv &  \int {\mathcal D}h^{\scriptscriptstyle (2)} e^{- \beta H_\mathrm{eff}^{\scriptscriptstyle \mathrm{(1,2)}}}  \\
      & = &    \int_{- \infty}^{\infty}  \!\!\!\!\! \d a_\vec{0}^{\scriptscriptstyle (2)}  
           \bigg( \prod\nolimits_{\vec{q}} 
          \int_{- \infty}^{\infty}  \!\!\!\!\! \d a_\vec{q}^{\scriptscriptstyle (2)}  
          \int_{- \infty}^{\infty}  \!\!\!\!\! \d b_\vec{q}^{\scriptscriptstyle (2)}  \bigg) 
          e^{- \beta H_\mathrm{eff}^{\scriptscriptstyle \mathrm{(1,2)}}}    ,  \hspace{0.82cm}
\end{eqnarray}
what leads to an effective Hamiltonian for the upper membrane, 
\begin{eqnarray} 
  H_\mathrm{eff}^{\scriptscriptstyle (1)}    
  & = &  L_x L_y \Big( \kappa_1 |\vec{q}|^4 + \gamma_1   - \frac{\gamma_1^2}{\kappa_2 |\vec{q}|^4 + \gamma_1 + \gamma_2  
                   - \frac{\gamma_2^2}{\kappa_3 |\vec{q}|^4 + \gamma_2}    } \Big) \,   \nonumber \\
  &&     \hspace{-0.25cm}   \times     \big( {a_\vec{q}^{\scriptscriptstyle (1)}}^2  + {b_\vec{q}^{\scriptscriptstyle (1)}}^2 \big)  
  \; + \frac{1}{2} L_x L_y \big( \gamma_1 - \frac{\gamma_1^2}{\gamma_1 + \gamma_2}) \, {a_\vec{0}^{\scriptscriptstyle (1)}}^2    .
\end{eqnarray} 

Consequently, the fluctuation spectrum of the upper membrane for non-vanishing wave vectors is 
(cf.\ equation (\ref{eqn:spectrum_2co_not0})) 
\begin{eqnarray}  \label{eqn:3M_co_top}
   \langle \, |h^{\scriptscriptstyle (1)}_{\vec{k} \neq \vec{0}}|^2 \, \rangle 
   & = &  \frac{k_B T / L_x L_y}{ \kappa_1  k^4 + \gamma_1   
                  - \frac{\gamma_1^2}{\kappa_2  k^4 + \gamma_1 + \gamma_2 
                                      - \frac{\gamma_2^2}{\kappa_3  k^4 + \gamma_2} } } \,  . ~~~~~
\end{eqnarray} 
For the translational mode ($\vec{k} = \vec{0}$) this expression would diverge, but the exclusion of 
$a_\vec{0}^{\scriptscriptstyle (3)}$ 
ensures the spectrum to be finite at the origin, 
\begin{eqnarray} 
   \langle \, |h^{\scriptscriptstyle (1)}_{\vec{0}}|^2 \, \rangle 
   & = &  \frac{k_B T / L_x L_y}{\gamma_1 - \frac{\gamma_1^2}{\gamma_1 + \gamma_2}  }  
   \:= \:   \frac{k_B T / L_x L_y}{\frac{1}{\frac{1}{\gamma_1} + \frac{1}{\gamma_2}  }  }    \ ,
\end{eqnarray} 
where the harmonic average of the two spring parameters enters (like for two springs in series).

Now we consider the middle membrane ($i=2$) and integrate out the independent variables belonging to the upper membrane ($i=1$) 
in equation (\ref{eqn:Def_H_1,2}), 
\begin{eqnarray}  
   e^{-\beta H_\mathrm{eff}^{\scriptscriptstyle (2)}} 
      & \equiv &  \int {\mathcal D}h^{\scriptscriptstyle (1)} e^{- \beta H_\mathrm{eff}^{\scriptscriptstyle \mathrm{(1,2)}}}  \\
      & = &    \int_{- \infty}^{\infty}  \!\!\!\!\! \d a_\vec{0}^{\scriptscriptstyle (1)}
           \bigg( \!\! \prod\nolimits_{\vec{q}} 
          \int_{- \infty}^{\infty}  \!\!\!\!\! \d a_\vec{q}^{\scriptscriptstyle (1)}  
          \int_{- \infty}^{\infty}  \!\!\!\!\! \d b_\vec{q}^{\scriptscriptstyle (1)}  \!\! \bigg) 
          e^{- \beta H_\mathrm{eff}^{\scriptscriptstyle \mathrm{(1,2)}}}    ,     \hspace{0.95cm}
\end{eqnarray}
what leads to an effective Hamiltonian for the middle membrane, 
\begin{eqnarray} 
  H_\mathrm{eff}^{\scriptscriptstyle (2)}
  & = &  L_x L_y \bigg( \kappa_2 |\vec{q}|^4 + \gamma_1 - \frac{\gamma_1^2}{\kappa_1 |\vec{q}|^4 + \gamma_1}    + \gamma_2    \\
      &&                                        - \frac{\gamma_2^2}{\kappa_3 |\vec{q}|^4 + \gamma_2} 
                        \bigg) \,  
           \times     \Big( {a_\vec{q}^{\scriptscriptstyle (2)}}^2  + {b_\vec{q}^{\scriptscriptstyle (1)}}^2 \Big)  
   + \frac{1}{2} L_x L_y \gamma_2  {a_\vec{0}^{\scriptscriptstyle (2)}}^2    \,  .   \nonumber
\end{eqnarray} 

For non-vanishing wave vectors the corresponding spectrum of the middle membrane reads 
\begin{eqnarray} \label{eqn:3M_co_middle}
   \langle \, |h^{\scriptscriptstyle (2)}_{\vec{k} \neq \vec{0}}|^2 \, \rangle 
   & = &  \frac{k_B T / L_x L_y}{ \kappa_1 k^4  
                  + \frac{1}{  \frac{1}{\kappa_1  k^4} + \frac{1}{\gamma_1} } 
                  + \frac{1}{  \frac{1}{\kappa_3  k^4} + \frac{1}{\gamma_2} } }    \nonumber \\ 
   & \equiv &  \frac{k_B T / L_x L_y}{ \kappa_2 \, k^4 +  \gamma^{\scriptscriptstyle (2)}_{\mathrm{eff}}(k) }  \  .
\end{eqnarray} 
The expression for the effective spring parameter $\gamma^{\scriptscriptstyle (2)}_{\mathrm{eff}}$ is analogous to 
the effective spring constant of a four spring system, where two pairs are connected in series and these pairs 
are connected in parallel. 
The excluded $a_\vec{0}^{\scriptscriptstyle (3)}$ hinders the spectrum at the origin from diverging, 
\begin{eqnarray} 
   \langle \, |h^{\scriptscriptstyle (2)}_{\vec{0}}|^2 \, \rangle 
   & = &  \frac{k_B T / L_x L_y}{\gamma_2}    \ ,
\end{eqnarray} 
which is independent from the upper membrane.

\end{document}